\documentclass[fleqn,usenatbib]{mnras}
\usepackage[T1]{fontenc}
\usepackage{graphicx}	% Including figure files
\usepackage{amsmath}	% Advanced maths commands
\usepackage{amssymb}	% Extra maths symbols
\usepackage{txfonts}
\usepackage{enumerate}
\usepackage{bbding} % checkmark symbol

\usepackage{xcolor}
\usepackage{rotating}
\newcommand{\Msun}{M$_{\sun}$}
\bibpunct[,]{(}{)}{;}{a}{,}{,}

\usepackage{upgreek}
\newcommand{\orcid}[1]{\href{https://orcid.org/#1}{\includegraphics[width=10pt]{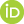}}}

\begin{document}

\title[Variations of $\alpha$-elements over Fe abundances in MWG]{Chemical evolution models: the role of type Ia supernovae in the $\alpha$-elements over Iron relative abundances and their variations in time and space}
\author[Cavichia et al.]{O. Cavichia$^{1}\orcid{0000-0002-7103-8036}$\thanks{cavichia@unifei.edu.br}, M. Moll{\'a}$^{2}\orcid{0000-0003-0817-581X}$, 
J.J. Baz{\'a}n$^{2}\orcid{0000-0001-7699-3983}$, 
A. Castrillo$^{3}\orcid{0000-0001-7743-5449}$, 
L. Galbany$^{4,5}\orcid{0000-0002-1296-6887}$,
%\newauthor
\and
I. Mill{\'a}n-Irigoyen$^{2}\orcid{0000-0003-4115-5140}$,
Y. Ascasibar$^{3,6}\orcid{0000-0003-1577-2479}$,
A.I. D{\'\i}az$^{3,6}\orcid{0000-0001-8529-7357}$,
H. Monteiro $^{1}\orcid{0000-0002-0596-9115}$
\\
$^{1}$ Instituto de F{\'i}sica e Qu{\'i}mica, Universidade Federal de Itajub{\'a}, Av. BPS, 1303, 37500-903, Itajub{\'a}-MG, Brazil\\
$^{2}$ Departamento de Investigaci\'{o}n B\'{a}sica, CIEMAT, Avda. Complutense 40. E-28040 Madrid. Spain.\\
$^{3}$ Departamento de F{\'\i}sica Te{\'o}rica, Universidad Aut{\'o}noma de Madrid,  E-28049, Cantoblanco (Madrid), Spain\\
$^{4}$ Institute of Space Sciences (ICE, CSIC), Campus UAB, Carrer de Can Magrans, s/n, E-08193 Barcelona, Spain. \\
$^{5}$ Institut d'Estudis Espacials de Catalunya (IEEC), E-08034 Barcelona, Spain.\\
$^{6}$ CIAFF–UAM. Centro de Investigaci{\'o}n Avanzada en F{\'\i}sica Fundamental, Spain\\
}

\date{Accepted Received ; in original form }
\pagerange{\pageref{firstpage}--\pageref{lastpage}} \pubyear{2023}

\maketitle
\label{firstpage}

\begin{abstract}
The role of type Ia supernovae (SN Ia), mainly the Delay Time Distributions (DTDs) determined by the binary systems, and the yields of elements created by different explosion mechanisms, are studied by using the {\sc MulChem} chemical evolution model applied to our Galaxy. We explored 15 DTDs and 12 tables of elemental yields produced by different SN Ia explosion mechanisms, doing a total of 180 models. Chemical abundances for $\alpha$-elements (O, Mg, Si, Ca) and Fe derived from these models are compared with recent solar region observational data of $\alpha$-elements over Fe relative abundances, [X/Fe], as a function of [Fe/H] and age. A multidimensional maximum likelihood analysis shows that 52 models are able to fit all these datasets simultaneously, considering the 1$\sigma$ confidence level. The combination of STROLG1 DTD and LN20181 SN Ia yields provides the best fit. The exponential model with very prompt events is a possible DTD, but a combination of several channels is more probable.
The SN Ia yields that include MCh or Near MCh correspond to 39 (75\%) of the 52 best models.
Regarding the DTD, 31 (60\%) of the 52 most probable models correspond to the SD scenario, while the remaining 21 (40\%) are based on the DD scenario.
Our results also show that the relatively large dispersion of the observational data may be explained by the stellar migration from other radial regions, and/or perhaps a combination of DTDs and explosion channels.
\end{abstract}

\begin{keywords}
Stars: supernovae: general --- Galaxy: abundances --- Galaxy: evolution ---Galaxy: disc ----Galaxy: halo ---
\end{keywords}

\section{Introduction}
\label{sec:intro}
Supernovae (SNe) explosions are one of the key ingredients in chemical evolution models, since they produce most of elements heavier than N. However, each SN type produces different proportions of elements. 
As an example, a typical type Ia supernova (SN Ia) produces around 0.01\,\Msun\ of Ca \citep{sei13a}, while a core collapse (CC) SN progenitor produces between 0.05 to 0.10\,\Msun\ \citep{Nomoto2013}. Massive stars, through CC SNe, pollute the ISM both with $\alpha$-elements, as e.g. O, Mg, Si, S, Ca \citep{Weinberg2017}, produced during the stellar evolution, and also iron peak elements, such as Cr, Mn, Fe, Co and Ni, created in the CC SNe explosion. Instead, Type Ia SNe mainly pollutes the ISM with the latter, leading to a decrease of the [$\alpha$/Fe] with increasing metallicity when they appear.

The importance of SNe Ia in the chemical evolution of galaxies resides in this amount of Fe released in these events. The mean Fe yield for a SN Ia is $y_{\rm Fe,Ia} \sim 0.7$\,\Msun\ \citep{iwa99,Mazzali2007,Howell2009}. Recent works give values in the range $[0.622-0.790]$\,\Msun\ per event (see Section \ref{sec:sn-yields}), 
while for a typical CC SNe, it is a factor of ten lower. When considering the percentage of explosions of all types of CC SNe (II, Ib, Ic, ..), \citep{Li2011b, Graur2017a, Graur2017b, Shivvers2017}, an average yield $<y_{\rm Fe,CC}> = 0.066$\,\Msun\ is obtained. 

Besides the yield of each individual SNe, it is necessary to consider their rates of production: there are between 7 and 8 times more CC SNe than SNe Ia, given their Hubble-time integrated production efficiency  $N/M_{*}$, i.e., the expected SN number by solar mass formed.
For CC SNe, this number is $N_{CC}/M_{*} \sim 0.01$\,\Msun$^{-1}$, depending on the assumed Initial Mass Function (IMF), while for SNe Ia it is in the range $N_{Ia}/M_{*} \sim [0.0016-0.054]$\,\Msun$^{-1}$, with field galaxies displaying the lowest value \citep{M17}. This way, SNe Ia and CC SNe contribute approximately to half of the Fe that is formed in a typical galaxy.

The stellar abundances may reflect this production of Fe through the relative abundance [$\alpha$/Fe], which retains the CC SN and massive stars production {\sl plus} the contribution of SNe Ia. In other words, for any $\alpha$-element as O, Mg, Si, S, Ca, or Ne, the relative abundances [$\alpha$/Fe]  depend on the number of SNe Ia that have exploded compared with the number of CC SNe. The ratio [$\alpha$/Fe] is, therefore, an indicator of the evolution of the region, mostly how the different SN rates have varied over time. If CC SN events were the only contributor to the Fe yield, there would exist a plateau in approximately [$\alpha$/Fe] $\sim 0.3$-- 0.5 \citep{M17}, defined by the corresponding yields of $\alpha$-elements and Fe from massive stars and CC SNe. In fact, this is the observed value for the metal-poorest stars of the Milky Way Galaxy (MWG). 

In turn, SN Ia progenitors are expected to reside in binary stellar systems, involving at least one white dwarf (WD) that accretes mass from a companion star increasing its mass, which approaches to the Chandrasekhar mass (MCh). The WD will explode after the evolution of the most massive component; that is, there is a {\sl delay} time after the creation of the binary system before the Fe created in that SNe Ia is ejected into the interstellar medium (ISM). This delay depends on the scenario involving the evolution of the binary system to reach the explosion. The final consequence is a decreasing in  [$\alpha$/Fe] from the initial ratio 0.3-0.5 from massive stars, once the Fe from SN Ia appears in the ISM. Thus, the chemical evolution may give clues about the proportions of both types of contributors.

The time distributions of SN rates are evaluated by using the so-called Delay Time Distribution (DTD) function, which is the expected SN rate (SNR) after an instantaneous star-formation burst or simple stellar population (SSP). A minimum time of delay is necessary to allow the most massive star to evolve, which would be a few tens of Myr for a star of 8\,\Msun. Then, the DTD must be modulated by the star-formation rate (SFR) of each region or galaxy in order to compute the total SNR within it. In recent years, improved estimates for DTDs of different SN species (CC and Ia) have been available. There are different methods to empirically infer the SN Ia DTD  \citep[see][ for a review]{Maoz2014} from the existing data. From the theoretical side, the SN Ia DTD was computed some years ago using binary star population synthesis (BPS) models \citep{Yungelson2000, Greggio2005}. These simulations computed the DTD of the different scenarios by using probability distribution functions for stellar orbital parameters and following the mass transfer in the system.

There are important differences in the evolution after the birth of the binary system until the explosion, depending on the assumed scenarios for that system. The two common ones are the single degenerate (SD) systems, where a single WD accretes mass from a non-WD star, such as a main sequence star, a red super giant (RSG), or a He-rich star; and the double degenerate (DD) systems, where both stars are WD, with at least one CO WD.

For the DD case, the time to reach the merger strongly depends on the separation of the WDs in the post-common envelope phase \citep{Totani2008}. The BPS models give for this case a DTD~$\propto t^{-1}$ \citep{Katz2012, Toonen2012} after a certain delay. 
There exist models showing a great variety of results depending on their parameters, as e.g. the exponential model \citep{strolger} but, in general, they predict a time dependence compatible with a power law, even with short delays \citep{Meng2010, Mennekens2010, Hachisu2008, Bours2013}. For times shorter than 30 -- 40\,Myr, which is the mean lifetime of stars with 8\,\Msun, we do not expect any SN Ia event, since this is the necessary time for the binary system, assuming it has a total mass in the range [3 -- 16]\,\Msun, evolves.

For the SD scenario, the mass of the non-degenerate donor star was very restricted in the early models, only 2--3\,\Msun, by the accretion conditions \citep{Langer2000}.  Recent studies, however, include different interactions in their BPS codes such as Roche-lobe overflows or wind ejecta interactions, expanding the mass range up to 8\,\Msun\ stars and some red giant stars. Depending on this mass, the resulting delay may be as short as $\sim$ 100\,Myr or even 40\,Myr \citep{Greggio2005}.

Other studies use different approaches to recover the DTD. Most of these observational DTDs find a similar dependence $\propto t^{-1}$ in a continuous way with no cut-off times, which supports the DD scenario. Sometimes a prompt population of the order of the 50\% of SNe Ia taking place with delay times shorter than 500\,Myr is found, which seems to indicate the SD scenario as probable. There exist, however, some tension about the existence or not of such short delays, without a consensus about the dominant channel for the SNe Ia.

On the other hand, the nucleosynthetic yields produced by SNe Ia are dependent on the explosion mechanisms, and finally determined by the density of the matter at which the thermonuclear burning starts. According to this, the yields of elements ejected by SNe Ia are usually divided in two categories depending on the mass of the WD: some SNe Ia explode when the Chandrasekhar mass is reached --or almost reached-- (MCh explosions), but there also exists the possibility that the WD explodes before to reach that MCh, giving place to a sub-MChandrasekhar (sub-MCh) explosion. \footnote{Even there are some DD that, due to angular momentum loss, explode with masses larger than MCh, the so-called super-MCh cases \citep{Piersanti2003}.}

The relationship between the SN Ia formation scenarios (SD or DD) and the WD masses involved in the explosions (sub-MCh and MCh) is a complex question and a current open debate in astronomy (e.g \citealt{Ruiter2020,Pakmor2022,Liu2023}.
There is no clear consensus either about the SN Ia progenitor system and the actual explosion mechanisms, that is, the two unknown points: which is the companion star donating the mass to the WD, (and the process of mass transfer and its timescale) and which is the WD mass or density when the SN Ia explodes are still unclear. Therefore, this dichotomy SD/DD or sub-MCh/MCh has changed very much in the last years. The idea that SN Ia originating in a given scenario tend to be associated with WD reaching a certain mass, is beginning to be considered too simple. In practice, it is not always easy to determine the exact mass of a WD before its explosion as a SN Ia. 
Actually, the diversity of the observed SN Ia is large, as it is shown in Figure 1 from \citet{Taubenberger2017}, what makes it difficult to classify them in only two categories.

In last years, there have been a lot of works developed around the explosion mechanisms besides the classic near-MCh explosion of a WD in a SD binary system \citep{Gilfanov2010, Sim2013}, or the violent merger of two WDs in a DD systems \citep{Iben1984, Guillochon2010, Pakmor2013}, the two old basic channels.  The mass accretion process and conditions in binary systems can vary widely. In some SD systems, the WD can accumulate enough mass to reach MCh, while in other cases, it can explode before reaching that critical mass (see for instance the SD models for both sub-MCh and MCh from \citealt{Greggio2005}). In the DD scenario, although the WDs involved are expected to be less massive than MCh, not all mergers will result in sub-MCh explosions, as the details of the merger are complex and may vary. There are some studies that support that a large fraction of SNe Ia comes from this sub-MCh channel, either by SD or DD He mass transfer scenario or through DD violent mergers \citep{Ruiter2009,Gilfanov2010,Shappee2013,Goldstein2018, Kuuttila2019,Flors2020}. Calculations of DTD with a DD scenario for MCh are also available \citep{Ruiter2011}.

\citet[][\, see his Table 1]{Soker2019} claims that now there are until five or more different explosions channels, such as: 1) single degenerate with a time of delay until the explosion from the merger (SD with MED); 2) Core degenerate (CD); 3) Double degenerate (DD with MED); 4) Double detonation (DDet); and 5) WD-WD collision. The CD and DD-MED models explode with MCh, while DD, SD-MED and DDet do that with sub-MCh. In turn,
\citet[][\,see Table 1]{Ruiter2020} established that a SN Ia may take place by: a) a MCh WD that accretes mass in a SD scenario by a delayed detonation (or failed detonation) also named deflagration-to-denotation transition -DDT- model; b) a sub-MCh WD accreting helium-rich material that explodes by a double detonation in a SD scenario (DDet), through the He mass accretion from a companion \citep{Nomoto1982, Greggio2005, Woosley2011, Bildsten2007,Kromer2010, Ruiter2011}; or c) a double WD merger with MCh or sub-MCh through a delayed detonation or double detonation. \citet{NL2019} clarified that "The sub-MCh mass explosions could occur in both SD and DD detonations". \footnote{Besides these scenarios, there are others, as three-stars systems, with two CO-WD and other non-WD star, which merger by the perturbation of their orbit due to the existence of a third one, reducing the time for the explosion \citet{DiStefano2020}.}
In their recent review, \citet[][\, see their Fig.3]{Liu2023} give a detailed explanation of these mechanisms about the different channels, where the authors specify that it is possible to have sub-MCh explosions in both SD and DD scenarios, and there are new scenarios with either one or two degenerate WD that allow explosions with MCh or sub-MCh.
Several of these research groups have calculated different explosions with sophisticated burning and explosion simulations, obtaining the corresponding SN Ia yields \citep[][\, see details in Section \ref{sec:sn-yields}]{iwa99, Shen2018, leung2018, leung2020, Eitner2020, Gronow2020}.

Thus, our idea is to use the fact that the [$\alpha$/Fe] {\sl vs.} Age and/or  [$\alpha$/Fe] {\sl vs.} [Fe/H] relations are able to trace the chemical enrichment produced by SNe Ia along the evolution, seeking to distinguish among the described scenarios, in order to clarify which DTD is the most probable. Furthermore, we will vary the sets of yields for SN Ia coming from different authors by assuming different explosion mechanisms. Therefore, the basic objective of this work is to compute the same chemical evolution model for the MWG varying the DTD prescriptions as well as the SNe Ia yields, in order to, after comparing results with observations of the solar neighbourhood, to search for the best model, in particular which DTD and explosion mechanism are the most probable. We assume in this work that both inputs are independent, since most of works devoted to DTD or SN Ia yields give this information separately, by computing all possible combinations DTD$+$yields; however, we are conscious that some models/combinations could be less realistic than others. We will comment this problem when analysing the results.

The relative abundance [$\alpha$/Fe], sensitive to the SN rates and this way to the DTD convolved with the SFR, has been previously used in other chemical evolution models \citep{Tinsley1979, Matteucci1986, DeDonder2004,Matteucci2006, Kobayashi2009, Tsujimoto2012} in MWG or in other nearby galaxies \citep{Walcher2016}. As an example, \citet{Calura2007} find some evidence of SN Ia with short delays from the abundance--age relation. More recently,
\citet{palla2021} have contributed to this subject by computing chemical evolution models with a set of SN Ia yields from different authors and simulating the three more probable explosion mechanisms within a model with {\bf only one DTD}. They found that the classical W7+WDD models (2 channels) produce similar patterns that near-MCh and sub-MCh mass models. 

Our compilation of observational data is presented in Section \ref{obs}. Section \ref{model} describes our MWG chemical evolution model, in particular the different DTD prescriptions under study, discussed in Section \ref{dtd}, and the yields used for SNe Ia described in Section \ref{sec:sn-yields}. The results of the comparison model {\sl vs.} data are discussed in Section \ref{Results}. Our main conclusions are summarised in Section ~\ref{Conclusions}. Furthermore, we give as Supporting Material the normalisation process of the observational data sets in Appendix \ref{AppA}; and the calibration of the MWG model in Apppendix \ref{AppB}.

\begin{figure}
\centering
\includegraphics[width=0.5\textwidth]{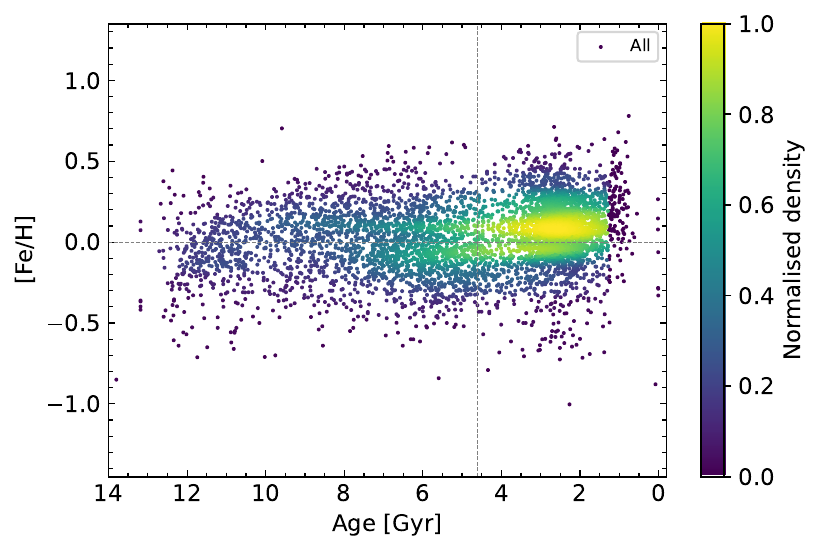}
\caption{Observed Iron abundances [Fe/H] as a function of the stellar age for 5622 stars from different authors as used in this work: GALAH-DR3 survey \citep{buder2021}, \citet{chen00} and HARPS-GTO \citep{bertran2015,suarez-andres2016,suarez-andres2017, dm17, dm19}, as described in Table~\ref{table:obs}). The density colour map is indicated by vertical colour bar. The dashed lines represent the solar values $\mbox{[Fe/H]}=0$ and age $\sim$ 4.6 Gyr.}
\label{feht_obs}
\end{figure}

\section{Observational data: stellar catalogues}
\label{obs}

Stellar data will be used to compare with the [X/Fe] relative abundance from models and to be able to determine which of them is the best one to fit these data, the most ones affected by variations in DTDs and SN Ia yields. Here, we describe these data sets.

To compare the model predictions for the time evolution of the disc, we searched for observations of stars with available elemental abundances for Fe and for $\alpha$-elements (as many of them as possible) and also age estimates. To conduct the statistical analysis, we only select datasets that include all the analysed chemical abundances and ages. This allows us to construct a complete sample, meaning that all the plots analysed in this work contain the same stars. We used here the stellar data from several authors/surveys, as given in Table~\ref{table:obs}, in particular those from \citet{chen00}, the HARPS-GTO and the GALAH-DR3 samples, which fulfil these conditions. This table describes the references for each dataset and the names adopted through this work to refer to the datasets.

\begin{figure*}
\centering
\includegraphics[width=\textwidth]{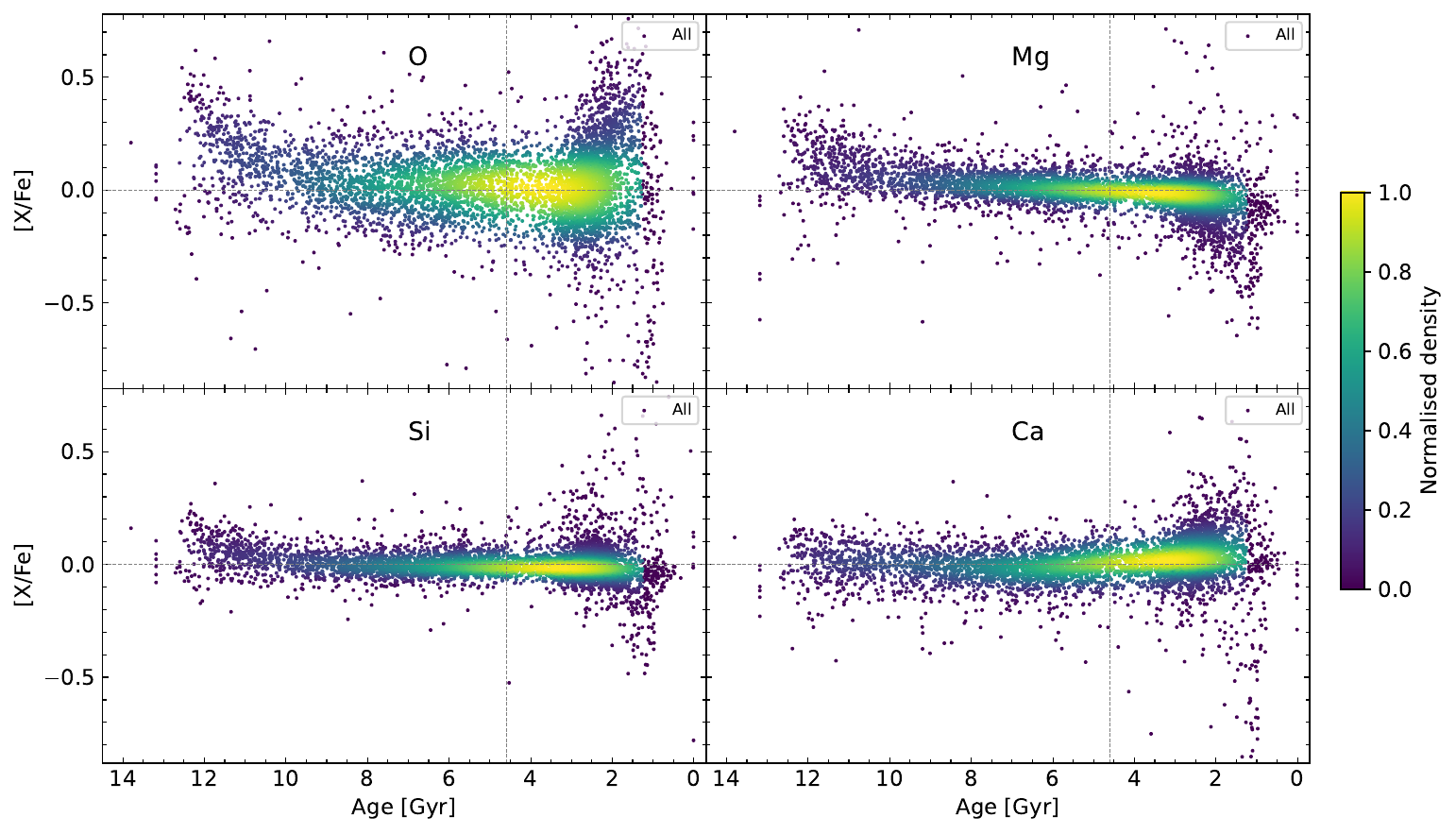}
\caption{Observed $\alpha$-elements abundances, [X/Fe], as a function of the stellar age. 
Data correspond to 5622 stars from all the datasets used in this work, as summarised in Table~\ref{table:obs}, for each element, where $X$ stands for O (upper left), Mg (upper right), Si (bottom left) and Ca (bottom right). The scale of the colour-density in each plot is indicated by the vertical colour bar at the right.}
\label{xfe_age_obs}
\end{figure*}

\citet{chen00} gathered high-resolution and high signal-to-noise (S/N) spectra for a sample of 90 F and G main-sequence disc stars. The observations were performed with the Coud\'{e} Echelle Spectrograph attached to the 2.16 m telescope at Beijing Astronomical Observatory (Xinglong, PR China). They provide metallicities in the range $-1.0 <$ [Fe/H] $< +0.1$, with approximately the same number of stars in each metallicity bin of 0.1\,dex. Stellar ages are also provided from interpolated evolutionary tracks. Their $\alpha$-chemical abundances for O, Mg, Si, Ca were used in our analysis.

\begin{figure*}
\centering
\includegraphics[width=\textwidth]{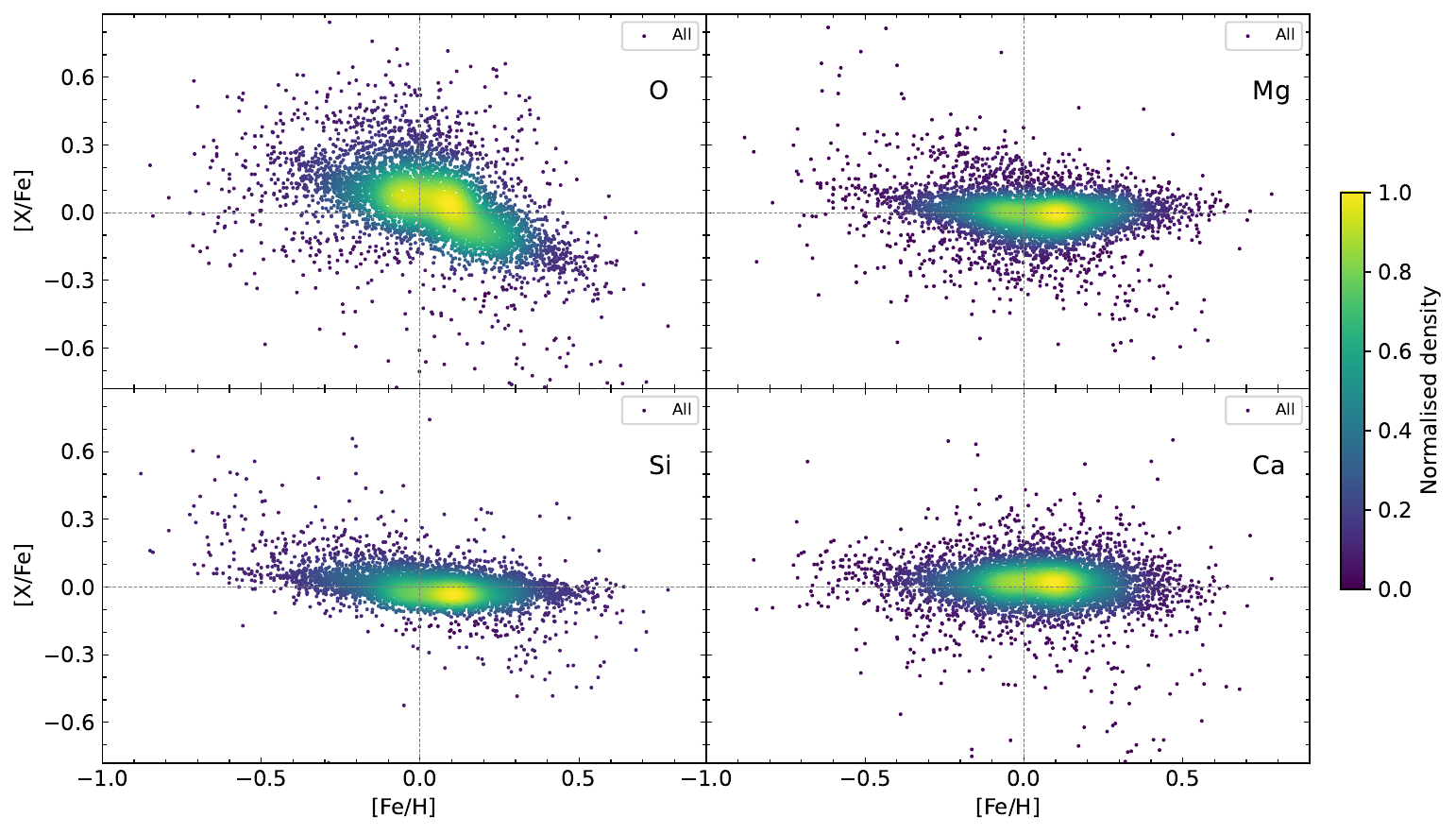}
\caption{The same as Fig. \ref{xfe_age_obs} but for the observed $\alpha$-elements abundances, [X/Fe], as a function of [Fe/H].}
\label{xfe_feh_obs}
\end{figure*}

The HARPS \citep[HARPS-GTO,][]{mayor03,locurto10,santos11} sample is composed by the stellar data given in  \citet{dm10,dm15,bertran2015,suarez-andres2016,suarez-andres2017,dm17} and \citet{dm19}, which are compilations from the HARPS planet search programme, HARPS-1, observed with the UVES spectrograph installed at the VLT/UT2 Kueyen telescope (Paranal Observatory, ESO, Chile) during several campaigns. We have used the catalogues\footnote{\url{https://cdsarc.cds.unistra.fr/viz-bin/cat/J/A+A/624/A78}} taken from the CDS-Strasbourg.
\citet{bertran2015} provide stellar parameters and abundances for 762 stars observed at high-resolution and high S/N. We prefer to use their oxygen abundances derived from \ion{O}{I}\,6158 \AA\, line, since these authors suggest that this line can be used as a reliable indicator of oxygen abundance in solar-type stars. We selected [Fe/H] abundances and stellar ages for 65 disc dwarf stars from \citet{suarez-andres2016}. In \citet{suarez-andres2017} the [C/H], [Fe/H] abundances and  stellar ages for 1058 FGK solar-type disc stars were selected. \citet{dm17,dm19} give abundances for Mg, Si, Ca. The HARPS data from these different authors are, in principle, compatible, since the abundances are calculated with the same methods and codes. Using all sets together, we will simulate the real dispersion of data due to different methods and instruments. Finally, in \citet{dm19} the stellar ages are obtained, allowing the analysis of the time evolution in the solar vicinity. Additionally, we apply a quality selection criteria by choosing stars with precision better than 0.2\,dex in the abundances and also those with the flag {\it thin} in the catalogues, to select only stars that pertain to the thin disc component.

We have also used the information from the second Gaia data release already included in the high-resolution Galactic Archaeology with HERMES (GALAH) spectroscopic survey \citep{buder2021}. This survey aims at to analyse the structure of the Galactic disc components using a large number of stars in the MWG by studying the kinematics and chemical abundances. The total number of stars from this DR3 is $\sim$ 588\,571. However, not all of them have  similar precision for stellar ages and abundances. We have selected stars with error in ages smaller than 15\%. As GALAH-DR3 also provides astrometric parameters from Gaia DR2 \citep{gaia2016,gaia2018}, we have also selected stars with height over the galactic plane $|z| < 0.2$\,kpc, and within the solar distance, with Galactocentric distances in the range $7 \le R_{\rm gal} \le 9$\,kpc, and we have restricted to good parallaxes with relative uncertainties lower than 10\%. The geometric estimated distances propagated from \citet{bailer-jones2018} were used and the heliocentric distances were converted to Galactocentric distances by adopting the Sun's distance to the Galactic centre $R_{0} = 8.1$\,kpc \citep{gravity2018}. These cuts implies errors in the ages for the oldest stars of $\sim 1.5$\,Gyr, although the average value is slightly higher than 0.5\,Gyr. We have also selected stars for which the abundances of the elements Fe, O, Mg, Si and Ca are available, and also limiting their errors to 0.1\,dex. This sub-sample, which consists of 5\,098 stars, is used through this work and referred as GALAH. More details about this sub-sample can be seen at the Appendix~\ref{AppA}.

All these data are given using different solar abundances, as shown in the Appendix \ref{AppA}, Supporting Information, Table~\ref{AppA:solar}. Therefore, in order to use all of them together, we have normalised the data to the same solar abundances scale from \citet{Lodders2019}. After this, differences in their scale could still be present, since we consider that all abundances [Fe/H] and [X/Fe] must be zero for ${\rm [Fe/H]=0}$ and at the Sun age of $4.6$\,Gyr (when the time since the Galaxy formation is $t=8.6$\,Gyr). That is, the data must be around the solar values at the solar metallicity and age. We have therefore shifted each cloud of data to locate the solar value in its centre. Further details about the normalisation process can be found in Appendix \ref{AppA}. In Table~\ref{table:obs} we provide the sets of data used and the chemical abundances and ages used for the analysis. The number of stars containing in each dataset is indicated in the second to last line and in the last line the references used for each dataset. The resulting combined data, considering all the datasets together, are represented in Fig.~\ref{feht_obs} for the age-metallicity relation and in Figs.~\ref{xfe_age_obs} and \ref{xfe_feh_obs} for the [X/Fe] relative abundances as a function of age and of [Fe/H], respectively.

\begin{table}
\caption{Set of catalogues and references of used data.}
\label{table:obs}
\resizebox{7cm}{!}{
\centering
\begin{tabular}{lccc}
\hline
 &  \multicolumn{3}{c}{Survey/Name} \\
\cline{2-4} \\ 
 & CHEN00 &  HARPS-GTO & GALAH\\ 
\hline 
[Fe/H] & \Checkmark & \Checkmark & \Checkmark \\
 
[O/H] & \Checkmark & \Checkmark & \Checkmark \\

[Mg/H] & \Checkmark & \Checkmark & \Checkmark \\

[Si/H] & \Checkmark & \Checkmark & \Checkmark \\

[Ca/H] & \Checkmark & \Checkmark & \Checkmark \\
Ages & \Checkmark & \Checkmark & \Checkmark \\
Stars & 78 & 446 & 5098\\
References & (1) & (2,3,4,5,6) & (7) \\
\hline
\end{tabular}
}
\\
\footnotesize{(1):\citet{chen00}; (2):\citet{bertran2015}; (3) \citet{suarez-andres2016}; (4): \citet{suarez-andres2017}; (5): \citet{dm17}; (6): \citet{dm19}; (7):  \citet{buder2021}}
\end{table}

\section{Summary description of galaxy evolutionary models}
\label{model}

Our {\sc MulChem} chemical evolution models, particularly the one applied to the MWG, have been described in \citet[][ hereinafter MOL15,MOL16,MOL17 and MOL19, respectively]{mol15,mol16,mol17,mol19}, updating the models from \citet{ferrini94,md05} for what refers to the inputs for modelling a given galaxy. We summarise here the changes done in this new version of the Galaxy model, related with stellar yields and SN Ia DTDs. The other model hypotheses and equations are widely described in the cited references. 

\subsection{Stellar yields.}
\label{yields}

To compute the elemental abundances, we use the technique based on the {\sl $Q$-matrix} formalism \citep{tal73,ferrini92,pcb98}. Each element $(i,j)$ of a matrix, $Q_{i,j}$ gives the proportion of a star which was initially element $j$, ejected as $i$ when the star dies. Thus,
\begin{eqnarray} 
Q_{i,j}(m)& =& \frac{m_{i,j,exp}}{m_{j}},\\
Q_{i,j}(m)X_{j}& = & \frac{m_{i,j,exp}}{m},
\end{eqnarray}
where $m_{i,j,exp}$ is the ejected mass of an element $i$ initially being other $j$. 
These numbers must be multiplied by the number of stars of mass $m$, given by the used IMF, here being the one from \citet[][ hereinafter KRO]{kro01} with limits $m_{\rm low}=0.15$\,M$_{\sun}$ and $m_{\rm up}=40$\,M$_{\sun}$ 

In this work, we use for the ejected mass from stars the stellar yield sets from \citet{cris11,cris15}, the FRUITY set, for low and intermediate mass stars, and from \citet{lim18} for massive stars. 
The first ones produced a first set of stellar yields for stars between 1 and 3\,\Msun\ \citep{cris11}, which is complemented with other set for masses in the range from 4 to 6\,\Msun\ \citep{cris15}. They follow carefully the evolution of each star, particularly the AGB phases with successive thermal pulses and dredge-ups. They give in their FRUITY web page (\url{http://fruity.oa-teramo.inaf.it/}) the complete information, stellar masses and core and surface abundances in each step, as well as the net and total yields for 400 isotopes and 10 metallicities. We have used their so-called {\sl total yields}, which actually are the total ejected mass (new and already existing in the star when formed) of each isotope to include in our code, and this way to obtain the Q's matrices for the eight stellar masses given by the authors. Moreover, using the information of the successive steps of the AGB stars evolution, that is, each remaining mass and the corresponding surface abundance, we have computed the secondary and primary components for $^{14}$N and $^{13}$C, by assuming that the production during the third dredge-up may be considered as primary and the rest is secondary. 
For massive stars, \citet{lim18} give the stellar yields for four metallicities ([Fe/H]=0, -1, -2, -3) and for nine stellar masses, from 13 to 120 \,\Msun\ (\url{http://orfeo.iaps.inaf.it}). They calculated the same sets for three different values of the stellar rotation velocity: 0, 150 and 300\,km\,s$^{-1}$, finding, in agreement with previous works  \citep{Meynet2006,chiappini2021}, that the rotation modifies the stellar yields, mainly the one for $^{14}$N, which appears as primary for low metallicities. In fact, since not all stars rotate at the same velocity, it is necessary to use a distribution of them for each metallicity. Following the recommendations of these authors, we have used the distribution shown in Fig.4 from \citet{Prantzos2018}. Therefore, we have finally a mix of stellar yields at each rotation velocity for each one of the 4 metallicities. 
Then, we have interpolated at the same 10 metallicities as FRUITY for low and intermediate mass stars,  to obtain a set of ejected masses for the whole mass range from 1 to 120 \,M$_{\sun}$ and 10 metallicities.
The stellar yields given by the authors for $^{24}$Mg, $^{28}$Si and $^{40}$Ca  have been multiplied by a factor of 2, 0.7 and 1.6, respectively, to better reproduce the observations. Finally, we added the stellar yields for SNe-Ia, see Section \ref{sec:sn-yields}. These new yields were already included in \citet{cavichia23}, successfully reproducing the Galactic bulge abundances ratios and their dependence along the evolutionary time. In this work we use {\sc starmatrix} \citep{starmatrix}, an Open Source Python code\footnote{\url{https://github.com/xuanxu/starmatrix}.} that compute the new ejected elements by each SSP, then included in the {\sc MulChem} model by a convolution with the SFR.

\subsection{Supernova rates and delay time distributions for SNe Ia}
\label{dtd}

We consider that all stars with $m > 8$\,\Msun\ and $m \le 40$\,\Msun\ will end their life as CC SNe (including all types together), although the limit of mass of stars that explode as CC SNe is under debate \citep{Doherty2017}. 
For a SSP, the total number of such events, $N_{CC}$, per unit stellar mass formed is determined by the IMF $\phi(m)$ as:
\begin{equation}
\frac{N_{CC}}{M_*} \simeq \int_{8\,M_{\sun}}^{m_{up}}\,\phi(m)\,dm,
\end{equation}
which gives a value of $\sim 0.0052$\,\Msun$^{-1}$ for a KRO IMF, defined within the range 0.15 and 40\,\Msun\ ($\sim 0.0054$\,\Msun$^{-1}$ if m$_{up}=100$\,\Msun), and 0.0083\,\Msun$^{-1}$ for a \citet{Salpeter1995} IMF, within the same mass range. The IMF from \citet{Chabrier2003} produces two times the number of the CC SNe created compared with a KRO IMF.

For an arbitrary star formation history, $\Psi(t)$, the instantaneous rate of CC SNe is given by:
\begin{equation}
\frac{dN_{CC}}{dt}(t) = \int_{8\,M_{\sun}}^{m_{up}} \Psi(t-\tau(m))\ \phi'(m)\ dm,
\end{equation} 
where $\phi'(m)$ denotes the adopted IMF subtracting the fraction of stars in binary systems that would yield SN Ia.

\begin{table*}
\centering
\caption{DTD functions used in the models computed in this work.}
\label{tab:dtdmodels}
\begin{tabular}{cccccc}
\hline
Num. & Name  & scenario & Delay/parameters & Reference & colour \\
\hline
1 & CASTRIL & DD &  50\,Myr & \citet{castrillo21} & sea green \\
2 & MAOZ017 & DD & 50\,Myr & \citet{M17} & light green \\
3 & CHE2021 & DD &  140\,Myr & \citet{chen2021} &  green\\
4 & GRCDD04 & Close DD & 400\,Myr&  \citet{Greggio2005} & dark green\\
5 & GRWDD04 &  Wide DD & 400\,Myr & \citet{Greggio2005} & orange\\
6 & STROLG5 & DD & $\upxi=-650$, $\omega=2200$, $\alpha=1100$ & \citet{strolger} & coral red \\
7 & GRCDD10 & Close DD  & 1\,Gyr & \citet{Greggio2005} & red \\
8 & GRWDD10 & Wide DD & 1\,Gyr & \citet{Greggio2005} & dark red\\
\hline 
9 & GRSDSCH & SD Mass sub-MCh &  40\,Myr & \citet{Greggio2005} & maroon\\
10 & GRESDCH & SD Mass Ch & 100\,Myr & \citet{Greggio2005} &  violet \\ 
11 & STROLG1 & SD & $\upxi=10$\,Myr, $\omega=600$, $\alpha=220$ & \citet{strolger} & lavender \\
12 & STROLG2 & SD & $\upxi=110$\,Myr, $\omega=1000$, $\alpha=2$  & \citet{strolger} & blue\\
13 & STROLG3 & SD & $\upxi=350$\,Myr, $\omega=1200$, $\alpha=20$ & \citet{strolger} & cyan \\
14 & STROLG4 & SD & $\upxi=6000$\,Myr, $\omega=6000$, $\alpha=-2$ & \citet{strolger} & light blue\\
\hline 
15 & RLPPC00 & combined & several Gaussians &\citet{RLP} & magenta \\
\hline
\end{tabular}
\end{table*}

In turn, the rate of SNe Ia explosions is given by the equation:
\begin{equation}
\label{def}
\frac{dN_{Ia}}{dt}(t) = \int_{\tau_{\rm min}}^{{\rm min}(t,\tau_{\rm max})} \frac{dN_{*}(t-\tau)}{dt}\ A_{Ia}\ DTD(\tau)\ d\tau,
\end{equation}
which involves the total number of stars 
\begin{equation}
\frac{dN_{*}}{dt}(t-\tau) = \Psi(t-\tau) \int_{m_{\rm low}}^{\rm m_{up}} \phi(m)\ dm,
\end{equation}
created per unit time at time $t-\tau$, the fraction of them ($A_{Ia}$) that eventually produce a SN Ia, and the delay time distribution $DTD(\tau)$ in terms of the age $\tau$.

If the IMF is normalised in mass, $\int_{m_{\rm low}}^{m_{\rm up}} m\ \phi(m)\ dm=1$, as usual, then $N_{*} = \int_{m_{\rm low}}^{m_{\rm up}} \phi(m)\ dm$ is the number of stars per unit stellar mass of a generation or single stellar population. This value depends again on the IMF, being 1.71~\Msun$^{-1}$ for the KRO IMF used here (2.02~\Msun$^{-1}$ for Salpeter). In that case, $N_{Ia}=N_{*}\times A_{Ia}$ gives the number of SN Ia produced by each stellar generation. Using observational data of SN Ia rates in galaxies, \citet{Greggio2005} found a value $A_{Ia}\sim 10^{-3}$, although this value may vary \citep{bonaparte13}.

This way, Eq.~\ref{def} may be written as:
\begin{equation}
\label{DTDeq}
\frac{dN_{Ia}}{dt} = N_{*}\times A_{Ia} \int_{\tau_{\rm min}}^{{\rm min}(t,\tau_{\rm max})}\Psi(t-\tau)\,DTD(\tau)\ dt.
\end{equation}

The function $DTD(\tau)$ is the Delay Time Distribution that describes how many type Ia SN progenitors die after a delay time $\tau$ for a SSP of mass 1\,\Msun. This function is usually normalised to 1 when integrated along the age of a galaxy
\footnote{This means that we may normalising  $\int_{\tau_i}^{\tau_x} N_{*}x A_{Ia}xDTD(\tau)\ d\tau= 1\times 10^{-3}$, when integrating along a time of 13.2\,Gyr.}:
\begin{equation}
    \int_{\tau_i}^{\tau_x} DTD(\tau)\ d\tau = 1,
\end{equation}
where $\tau_{i}$ is the minimum delay, or minimum lifetime for one star precursor of a SN Ia. Since it is usually assumed that the SNe Ia occur in binary systems where the mass limits are assumed to be between 3 and 16\,\Msun,  $\tau_i$ would correspond to the age of the most massive component possible, that is 8\,\Msun.
For the upper limit, we need to choose between the longer age corresponding to the lower mass that could exist in a binary system, $\sim$ 1.5\,\Msun\ and, if the conditions allow it, the time in which the explosion may occur. In practice, this $\tau_{x}$ may be as long as the Hubble time. 

\begin{figure}
\centering
\includegraphics[width=0.33\textwidth,angle=-90]{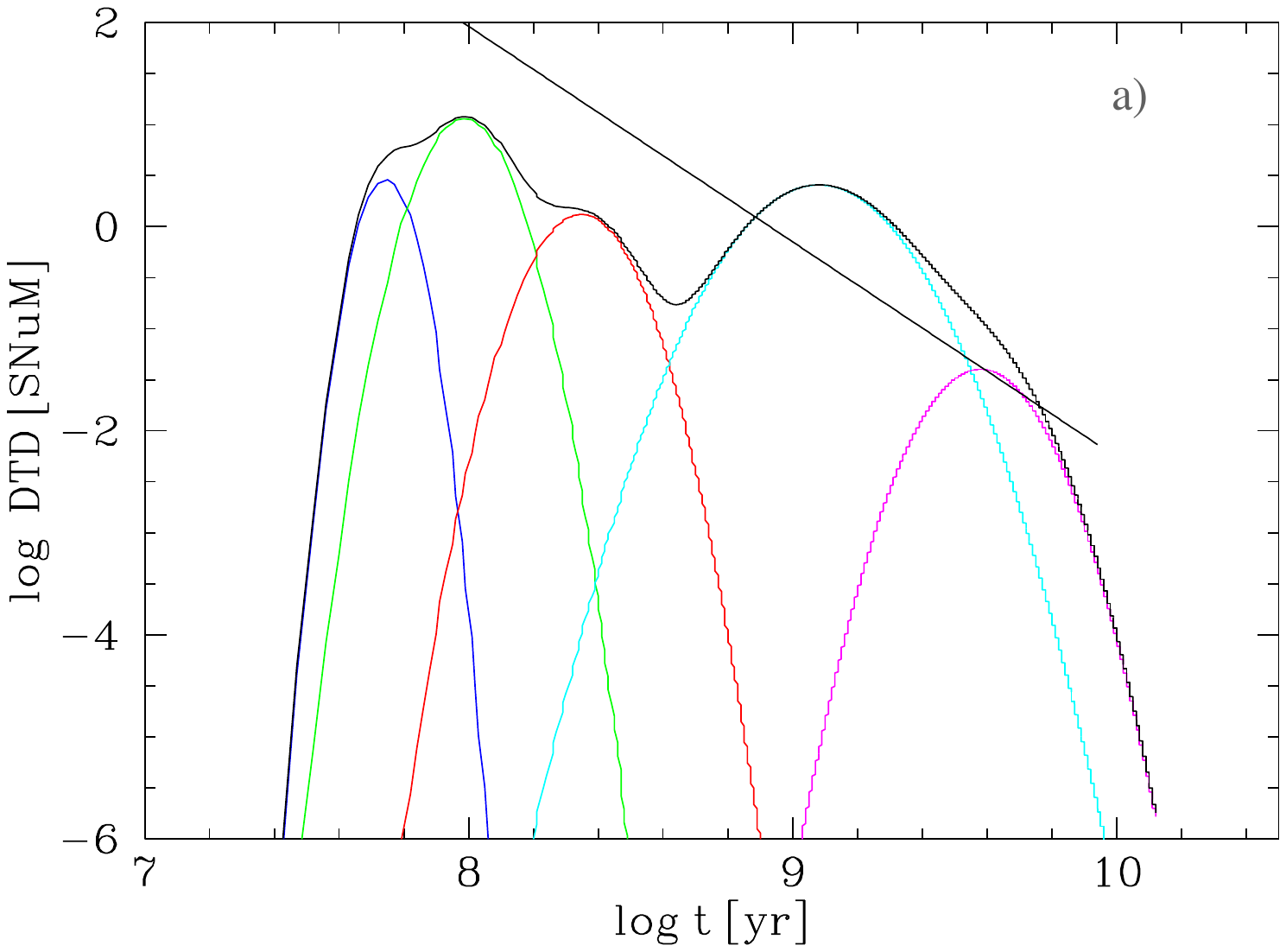}
\includegraphics[width=0.32\textwidth,angle=-90]{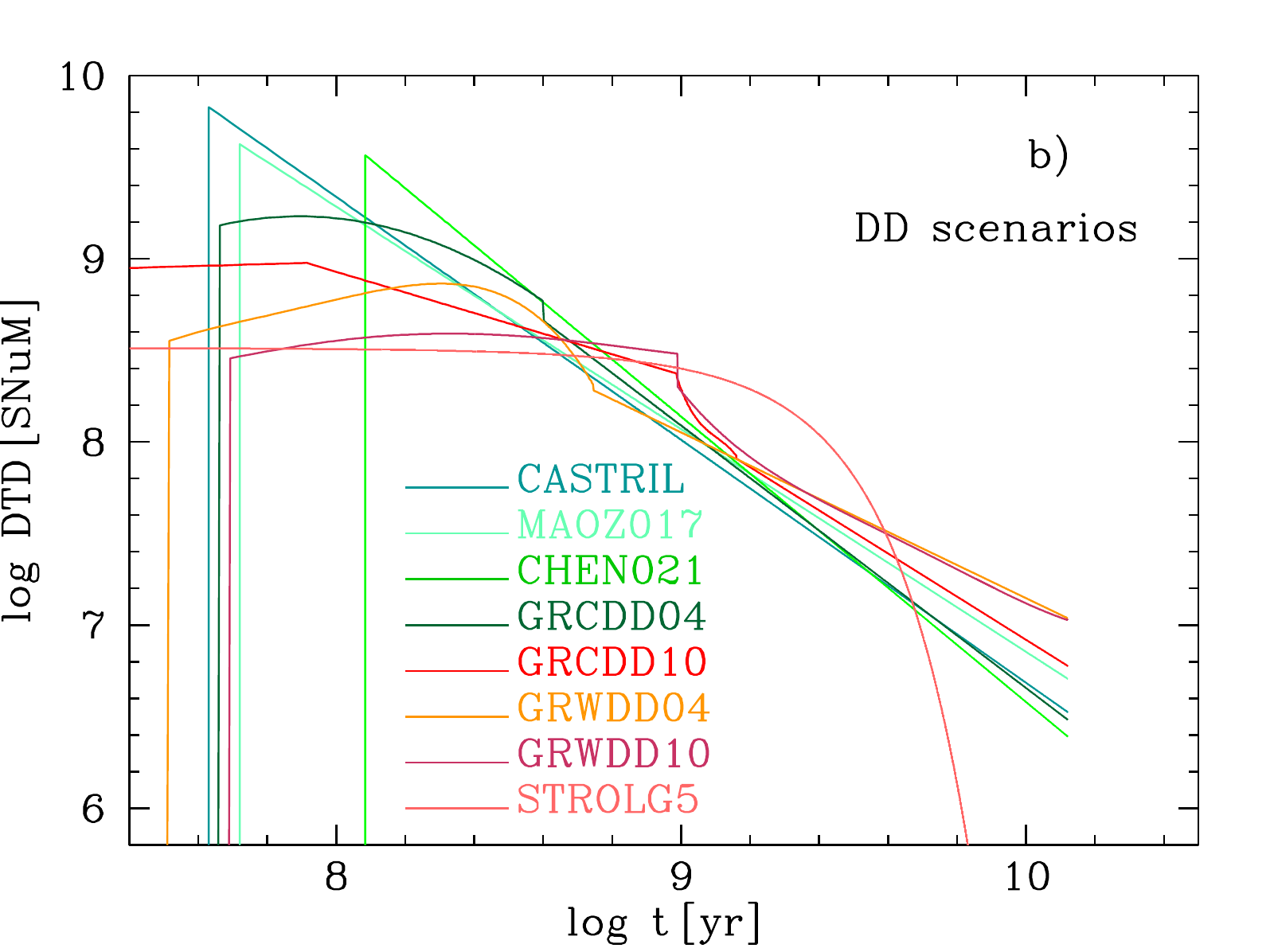}
\includegraphics[width=0.32\textwidth,angle=-90]{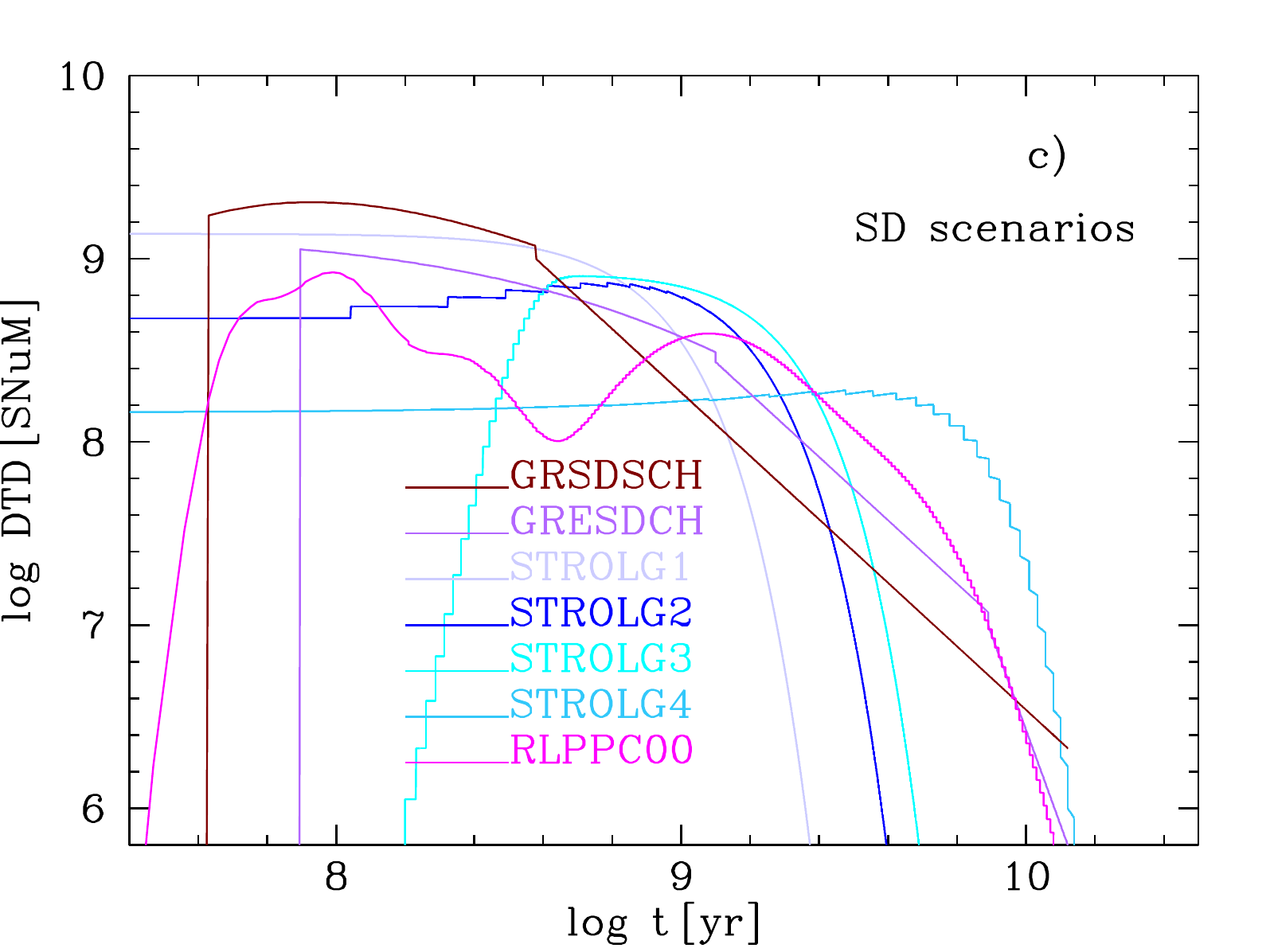}
\caption{Theoretical and empirical DTD functions:  a) the set of Gaussian curves, representing each one a different channels of SN Ia from RLPPC00, is shown as an example of combination of different channels. The straight line would correspond to a power law function of slope $\beta=-1$; b) for DD scenarios; c) for SD scenarios, as given by authors showed in Table~\ref{tab:dtdmodels}. RLPPC00 is also included in this panel although it represents a mix of scenarios. We have normalised all functions to give a value of 0.001\,SN\,M$_{\sun}^{-1}$ integrated along the life of the Universe. Each DTD is shown with a different colour as labelled. All SN rates are given in the usual unit of SNuM, number of SN that appear by 10$^{10}$\,M$_{\sun}$ in 100\,yr.}
\label{dtd_comp}
\end{figure}

Different progenitor scenarios for SNe Ia, and therefore different possible DTDs are at present supported by data. The {\sc MulChem} model in its original form \citep{ferrini92,ferrini94} used the technique explained in \citet{Greggio1983} to compute the SNe Ia rate from a secondary mass distribution. In more recent versions, from \citet{md05} until MOL19, the code incorporated the rates given by \citet{RLP97,RLP}, who provided a numerical table (private communication) with the time evolution of the SN rates for a SSP, computed under different assumptions or scenarios and their probabilities of occurrence. The final function, called RLPPC00, composed by several Gaussian functions, each one describing a given channel and showing a maximum at a different delay time, is shown in Fig.~\ref{dtd_comp}a).
In addition to the RLPPC00 DTD, which has been consistently used for the verification of our results, we have employed 14 additional DTDs, as listed in Table~\ref{tab:dtdmodels}. These were obtained for both DD and SD scenarios by various authors, following either theoretical \citep{Greggio2005} or observational prescriptions.

As explained in Section 1, there also exist some empirical works that deduce the SNR or DTD necessary to obtain the observed SN frequency. A review about this technique is given in \citet{maoz11}.
From an observational point of view, the first DTD studies from \citet{Mannucci2005, Mannucci2006} established a relationship between the SNR and the colour and Hubble type of the SN host galaxy, both measurements considered approximations to the SFR. After these first attempts, \citet{Sullivan2006, Li2011a, Li2011b, Smith2012} found two populations of SNe Ia: one associated to the SFR with short delays of 100--500\,Myr, the prompt population; and a tardy population associated to the galaxy mass and delays of up to 5\,Gyr. However, more recent studies have found a general relation between the SNR and the age of the underlying stellar population, implying a continuous DTD proportional to $t^{-1}$ \citep{Maoz2012b}, a result supported by theoretical models for DD scenarios.

Measuring the SNR in galaxy clusters, a DTD also proportional to $t^{-1}$, but with delays longer than 2\,Gyr up to 10\,Gyr, was found \citep{Barbary2012, Maoz2010}. The same result was obtained for field elliptical galaxies \citep{Totani2008}. \citet{Thomson2011, Raskin2009, Aubourg2008} also establish the existence of a prompt population that can dominate the SNR, even if the current SFR is low compared to the present value in the MWG \citep{Maoz2012b, Mannucci2009}. 
More recently some works computed the DTD from the volumetric rate evolution of SN Ia. Large SN surveys are then necessary to infer the SNR, while the SFR comes from cosmological simulations \citep{Graur2013, Rodney2014, Friedmann2018}, from the modelling of the colours for the galaxy \citep{Heringer2017, Heringer2019}, from the use of stellar population synthesis codes for individual galaxies \citep{Maoz2012a}, or even from integral field spectroscopy analysis of the host galaxies \citep{castrillo21}. 

This way, some power-law functions that vary with time from \citet{M17}, \citet{castrillo21}, and \citet{chen2021}, and are valid for the DD scenario, are used here:
\begin{eqnarray}
DTD_{\mathrm{MAOZ017}}&=& A\times t^{-1.1}, \mbox{ if } t> \tau=50\,\mbox{Myr} \nonumber,\\
DTD_{\mathrm{CASTRIL}}&=&B\times t^{-1.2}, \mbox{ if } t>\tau=40\,\mbox{Myr} \nonumber,\\
DTD_{\mathrm{CHE2021}}&=&C\times t^{-1.41}, \mbox{ if } t>\tau=120\,\mbox{Myr},
\label{Eq:DD-obs}
\end{eqnarray}
where A, B and C are normalisation constants. We have also added another function, STROLG5, as given by \citet[][ their figure 5, right panel]{strolger}, obtained with parameters given in that figure.
All these functions are represented in Fig.~\ref{dtd_comp}b, with different colour lines, as labelled.  In most of the empirical works, the exponent of the function is in the range $\sim -1.1$ to $-1.35$: $-1.1$ for MAOZ017, $-1.30$ for \citet{Friedmann2018}, $-1.34$ for \citet{Heringer2019} and $-1.1$ for CASTRIL; while the delay time $\Delta \tau$, as given in Eq.~\ref{Eq:DD-obs}, is more variable: 40\,Myr for CASTRIL, 50\,Myr for MAOZ017, and 120\,Myr for CHE2021. These latter authors established that although their results indicate a high confidence of $\Delta \tau= 120$\,Myr, and their slope is $-1$ consistent with the DD scenario, they cannot firmly discard other channels of SNe Ia with other times of appearance. 

Some theoretical functions from \citet{Greggio2005} are also used and included in Figure \ref{dtd_comp}b (GRCDD04 GRCDD10, GRWDD04 and GRWDD10), which correspond to the close and wide DD scenarios with time parameters 0.4 and 1.0\,Gyr. They are compatible with models represented by power law functions for times longer than some hundreds of Myr, except the one from STROLG5.

Moreover, we have also tried other DTDs that would correspond to the SD scenario. In that case, we have used the curves given by  \citet{Greggio2005} for her models SD with MCh and with sub-MCh (GRESDCH and GRSDSCH). To these theoretical prescriptions, we have added other 4 functions as given by  \citet[][ their Figure 5, left panel]{strolger}, obtained by modifying the parameters of their equation to fit the observational data from \citet{Nelemans2001}. All these functions and the one from RLPPC00, which is a mix of channels, are drawn Fig.~\ref{dtd_comp}c with different colour lines as labelled.

\begin{table*}
\centering
\caption{Set of yields for SNe Ia used in the models computed in this work.}
\label{tab:sniayields}
\resizebox{17cm}{!}{
\begin{tabular}{ccccccc}
\hline
Num. & Name  & type of explosion & model/table &reference & type of line \\
\hline
1 & IWA1999 & MCh & fast deflagration  W7/W70& \citet{iwa99} & solid\\
2 & SEI2013& MCh & DDT Table 2-N100 &\citet{sei13a} & light solid\\
3 & MO20181 & MCh & W7 DDT model Table 3 & \citet{mori2018} & dotted\\
4 & MO20182 & MCh & WDD2 DDT model Table 3 & \citet{mori2018} & light dotted\\
5 & LN20181 & Near MCh & DDT Table 3, low $\rho$ & \citet{leung2018} & short-dashed\\
6 & LN20182 &  Near MCh & DDT Table 4, medium $\rho$ & \citet{leung2018} & light short-dashed\\
7 & LN20183 & Near MCh & DDT Table 5, high $\rho$ & \citet{leung2018} & long-dashed \\
8 & BR20191 & MCh & DDT Table 3 & \citet{bravo2019} & dot-short-dashed\\
9 & BR20192 & sub-MCh & Table 4 & \citet{bravo2019} & light dot-short-dashed \\
10 & LEU2020 & sub-MCh & DDet Table 7 & \citet{leung2020} & short-long-dashed \\
11 & GR20211 & sub-MCh & DDet with He shell, Table A10, Table 3 & \citet{gronow2021} & dot-long dashed\\
12 & GR20212 & sub-MCh & DDet with He shell, Table A8, Table 4 & \citet{gronow2021} & light dot-long dashed\\
\hline
\end{tabular}
}
\end{table*}
Some years ago, \citet{Mennekens2010} performed synthesis of populations to produce models of SD and DD scenarios. They compare the resulting DTD curves with observational data \citep{Mannucci2005,Totani2008}, as seen in their Figure 2. They show that the SD scenario produces a maximum in 1.4\,Gyr and then disappears after 8\,Gyr. The DD scenario creates a curve proportional to $t^{-1}$ and it is the dominant channel. However, the SD alone does not fit the observed distributions, being DD or a mix of both channels SD$+$DD the most probable scenario. In fact, in that figure they presented a mix of both channels that reproduce well the data.  
Similarly, \citet{Ruiter2009,Ruiter2011} did similar models for 4 scenarios: DD, SD and HeR with MCh, and one double detonation sub-MCh scenario involving helium-rich donors. Computing the DTDs for the 4 cases and comparing with the observations, they show that the Sub-MCh may create both, a prompt ($<500$\,Myr) channel and a more delayed ($>500$\, Myr) one. 

Such as \citet{Nelemans2013} claimed, "the observed SN populations probably originated from a combination of multiple channels". That is, although the SNe Ia scenario defines mainly the delay time for the explosion will occur, a diversity of models is possible due to the different mass accretion rates, the existence or not of winds, the stripping rate of the companion, or the variations to radiate away the angular momentum, which change the time necessary to have a explosion in the DD scenario. Thus, theoretical models give delay times between 0.5 to 3\,Gyr for SD scenarios and a mean time larger than 0.3\,Gyr for DD scenarios. In fact, there is not a unique model able to reproduce satisfactorily the diversity of observations of SNe Ia. It is, therefore, quite probable that several channels are mixed to produce the SN Ia observed distribution \citep{Mernier2016, Hitomi2017}.
Although we have mainly used only one DTD in each model, except when using RLPPC00 that is by itself a mix of scenarios, it is necessary to take into account that the most probable solution will need two or more DTDs, either simultaneously or varying along the time, in the same model. 

\begin{figure*}
\centering
\includegraphics[width=0.3\textwidth,angle=-90]{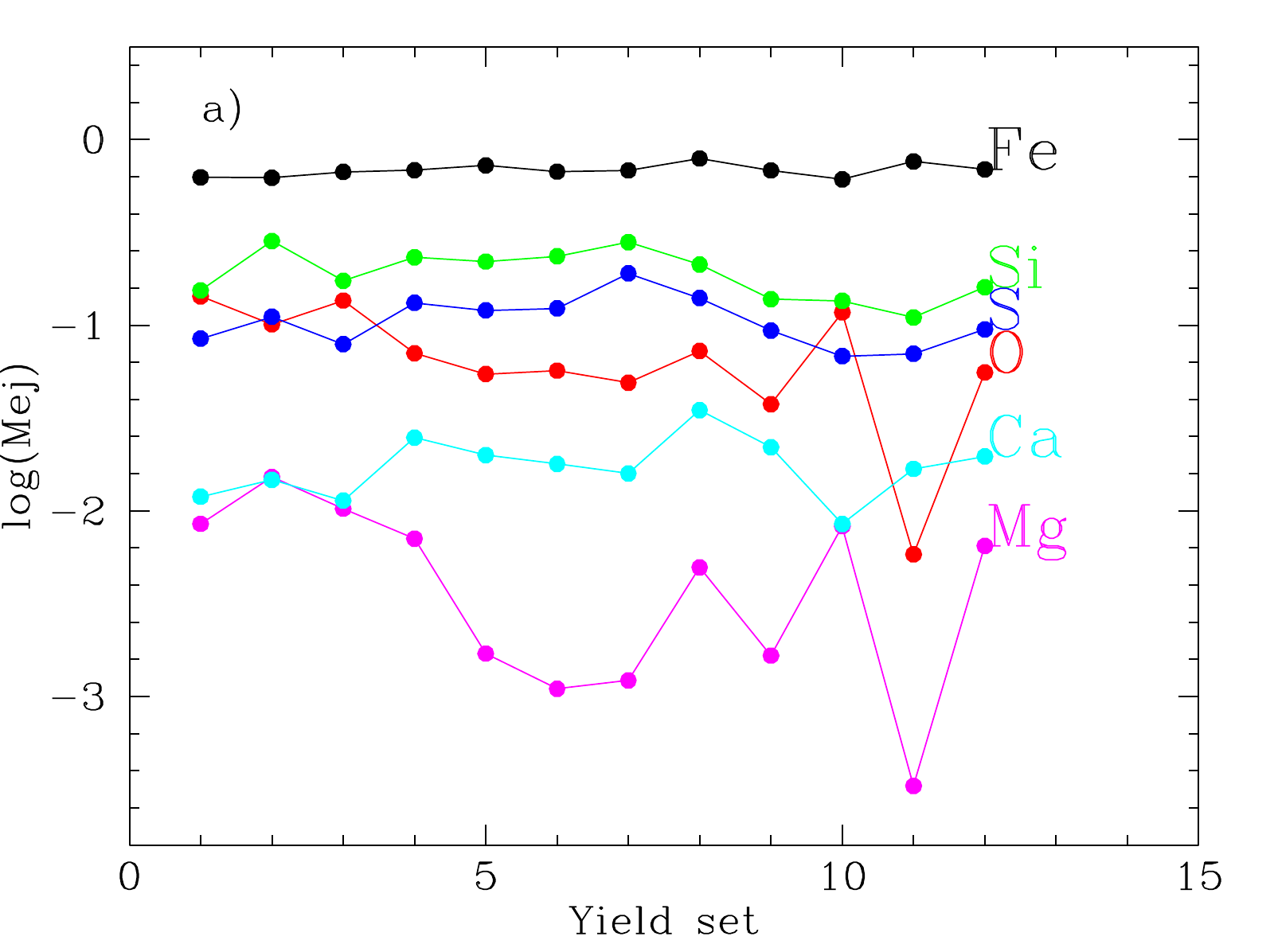}
\includegraphics[width=0.3\textwidth,angle=-90]{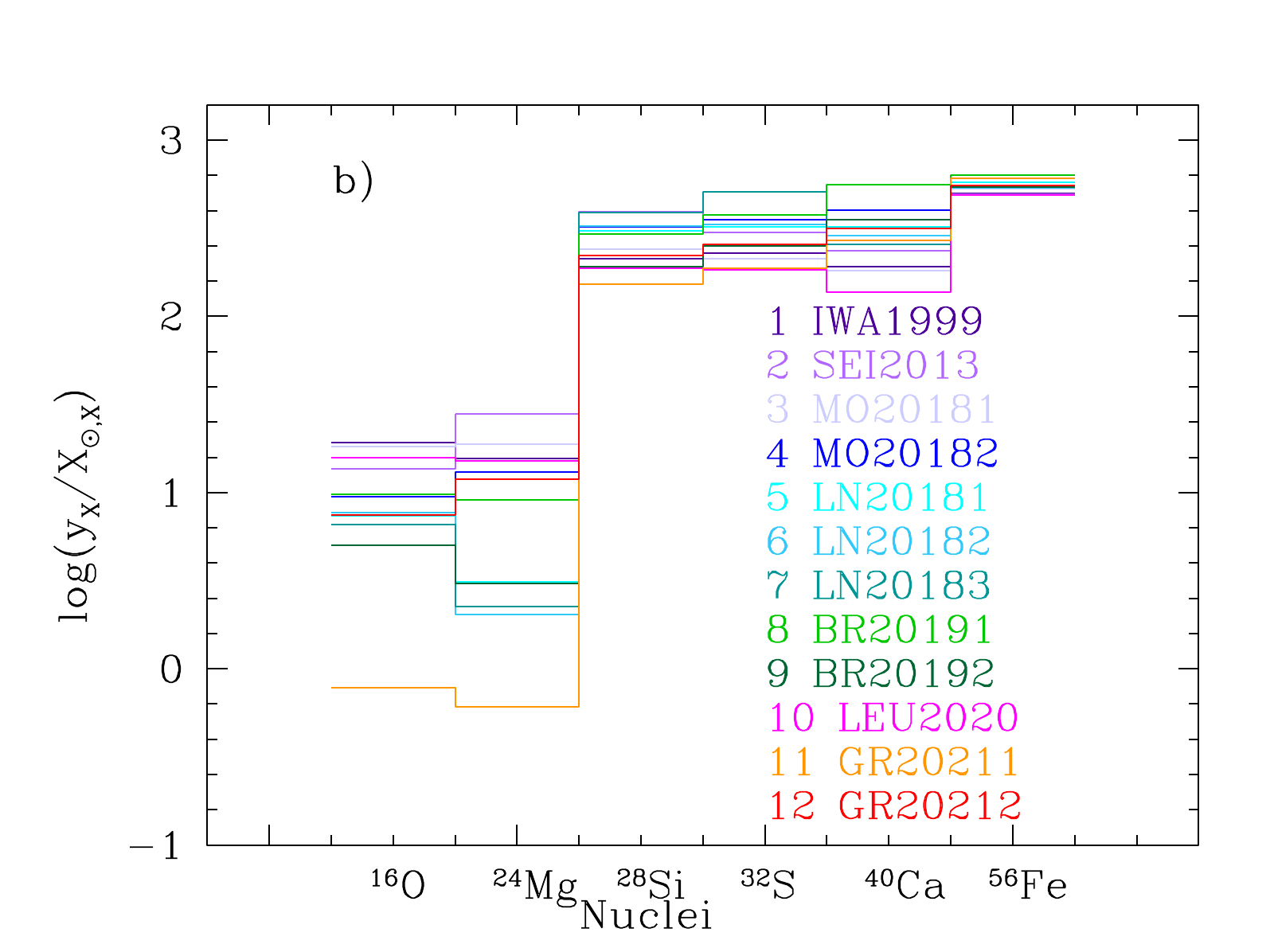}
\caption{Yields of different elements as ejected in the 12 different explosions models included in this work, numbered as given in Table~\ref{tab:sniayields}: a) yields given as the ejected mass  (logarithmic scale), $\log{M_{\mathrm ej}}$, from $^{16}$O to $^{40}$Ca plus $^{56}$Fe; b) zoom of the $\alpha$-elements yields from $^{16}$O to $^{40}$Ca and the one of $^{56}$Fe, given as the ratio of the ejected mass by solar abundances, $\log{y_{i}/X_{\sun}}$  from different authors as labelled.}
\label{fig:snia-yield}
\end{figure*}

\subsection{The SN Ia element yields}
\label{sec:sn-yields}

As explained above, there is a variety of possible thermonuclear explosion models, which produce elements in different quantities \citep{Arcones2023}. The explosion mechanisms may be due to different situations, which are widely explained in \citet[][ see their table 1 and Fig. 3]{Liu2023} and summarised as:
1) the burning of C in the centre of the star, when the WD reaches the MCh (the classical case): the simmering phase starts, with the consequent compression heating and a central burning that produces the thermonuclear runaway;
2) the burning of C in the outer envelope, outer ignition that occurs with a merger of two WD (DD scenario);
3) the burning of He, which starts in the surface of the star, when He is accreted from a companion rich in helium (He-WD or CO-WD with He envelope). This burning provokes a shock wave towards the centre of the WD and produces a WD detonation.

There exist differences in the described mechanisms, since the flame propagation may occur with different velocities: a) {\it detonation}, when the burning occurs at supersonic velocities; b) {\it deflagration}, when the burning occurs at subsonic velocity; and c) when a {\it deflagration to detonation transition} (DDT) occurs.

For this particular subsection we are interested in works where the nucleosynthesis yield sets are calculated for the different models. We refer to Table~\ref{tab:sniayields} for the details about the computed models in this work \citep{iwa99,sei13a,mori2018,leung2018,bravo2019,leung2020,gronow2021}. These models are divided into two categories: a) those where the explosion occurs at MCh (or near-Mch), and b) those where it occurs at sub-MCh.
The first ones are usually computed for a delayed detonation (DDT) channel in a SD or DD scenario; while the second ones are calculated for the double detonation (DDet) case in a SD scenario, or for the merging of two WD (violent mergers) in a DD scenario. Therefore, although {\sl a priori} there is no direct correlation between the SN Ia scenarios and the explosion mechanisms, there are some limitations: the SD scenario cannot produce an outer ignition, while triple systems occur basically with outer C ignition. The intermediate case core degenerate (CD) scenario does not occur with surface He-burning. But both scenarios SD and DD may occur with any mode of explosion. The key ingredient to obtain a SN Ia is the density and temperature in the burning time. This determines the ejected quantity of $^{56}$Ni. 

After the seminal work from \citet{Nomoto1982} with the classical W7 model, the revised model by \citet{iwa99} is the most frequently adopted in chemical evolution studies. The W7 model adopts a pure deflagration schema in a DD scenario and was conceived to reproduce the observational data, so that there are inherent physical limitations and additional 1D modelling limitations. The yields were calculated for solar metallicity and a 1/10 of the solar value. \citet{iwa99} also studied a slow delayed detonation with DDT and a channel CD. 

Later, \citet{sei13a} presented results for a suite of 3-dimensional, high-resolution hydrodynamical simulations of delayed-detonation explosions (DDT) giving four datasets for ${\rm [Fe/H]} = -2$, -1, -0.3 and 0. In this model, the detonation is delayed with three phases: 1) There is a deflagration with burning in a high density core; 2) The energy propagates with subsonic velocity and the star expands, so the density is now lower; 3) A detonation occurs after a time. \citet{sei13b} found that the yields show a great dependence on metallicity in both CC SN and SNe Ia, finding that the 50\% of SNe Ia would have explode as near-MCh WDs to reproduce the observations related to [Mn/Fe].

In recent years, there has been a growing number of studies that have computed models of explosions, providing the yields of elements generated by them. We have selected some of these works to utilise in our research \citep{leung2018, mori2018, gronow2021}. These studies employ various thermodynamic conditions, such as central densities, velocities, turbulent velocity fluctuations, and different strengths of the deflagration phase. Additionally, these models also allow for explosions with masses below the Chandrasekhar mass (sub-MCh). Also, there are two sets from \citet{bravo2019} giving yields able to reproduce the observations of SN Ia. In turn,
\citet{leung2020} have revised the W7 model for a MCh deflagration explosion with improvements on the nuclear reaction network in comparison with those from \citet{iwa99}, also included in our set of yields. 

The yields of different elements ejected used in this work correspond to twelve different sets obtained for different explosion mechanisms and given in Table~\ref{tab:sniayields}, where we define for each set a number (column 1) and an acronym or name of the yield set (column 2) that we will use later for identifying each one. The third column displays the type of explosion, the fourth column the chosen model or table, the fifth column the source reference, and the sixth column the line style to plot the models in the next figures. In some cases the authors give more than one table and we have selected one or more sets following their recommendations. All of them are plotted in Fig.~\ref{fig:snia-yield}. In panel a) we represent the ejected mass of each element as a function of the number of the yield set (column 1 of Table~\ref{tab:sniayields}); In panel b), we represent the abundance compared with the solar one from \citet{Lodders2019} as a function of the mass number A of each nuclei. Each set is drawn with a different colour as labelled. The yield for Fe, the most important element in this case, is essentially the same for all of them. For O, except GR20211, Si, and S, all tables give results within an order of magnitude. There are, however, important variations in the yields for $^{24}$Mg and $^{40}$Ca among the different sets, where models GR20211, LN20182 and BR20192 show the lowest yields values. 

\section{Results}
\label{Results}

We have computed 180 models for MWG with fifteen different possibilities of DTD functions, representing different scenarios for the binary systems and their evolution, as shown in Table~\ref{tab:dtdmodels}, combined with twelve different set of yields for different explosion mechanisms for the SNe Ia, as shown in Table~\ref{tab:sniayields}. As it is said in the Introduction, not all combinations of DTD-SN yields are compatible with our knowledge about SN Ia scenarios and channels. Thus, yields computed for delayed detonation (DDT) models with MCh explosions are compatible with DTD computed for SD scenarios, but the DDT explosion is also possible in the DD scenario. In turn, yields for sub-MCh are calculated for double detonation (DDet) channels in a SD scenario. The merging of two WDs as violent mergers may also occur at sub-MCh.  However, we have not used any set of yields valid for this case, and this way, yields sets for sub-MCh are here only valid for the DDet explosions in SD scenarios. The combinations sub-MCh yields $+$ DD DTDs that we have among the 180 computed models are not realistic in this work and these combinations must be interpreted with caution.

Our computations are based on the same model from MOL19, with the same total mass radial distribution, collapse time scales, and star formation efficiency for MWG. We have, however, updated the stellar yields to the those from \citet{cris11, cris15, lim18}, and we have also modified the stellar mean-lifetimes. For that reason, we have checked in Appendix~\ref{AppB}, Supporting Information, that models give, for the quantities not related with Fe or $\alpha$-element abundances, the same results than in previous works, reproducing equally well the MWG data: SFR and the time evolution of CNO abundances for the solar region, and the radial distributions of SFR, mass of gas and stars, and CNO abundances in the Galactic disc, all of them compiled in \citet[][ MOL15]{mol15}. 

\subsection{Supernova rates in the MWG}
\begin{figure}
\centering
\includegraphics[width=0.45\textwidth,angle=0]{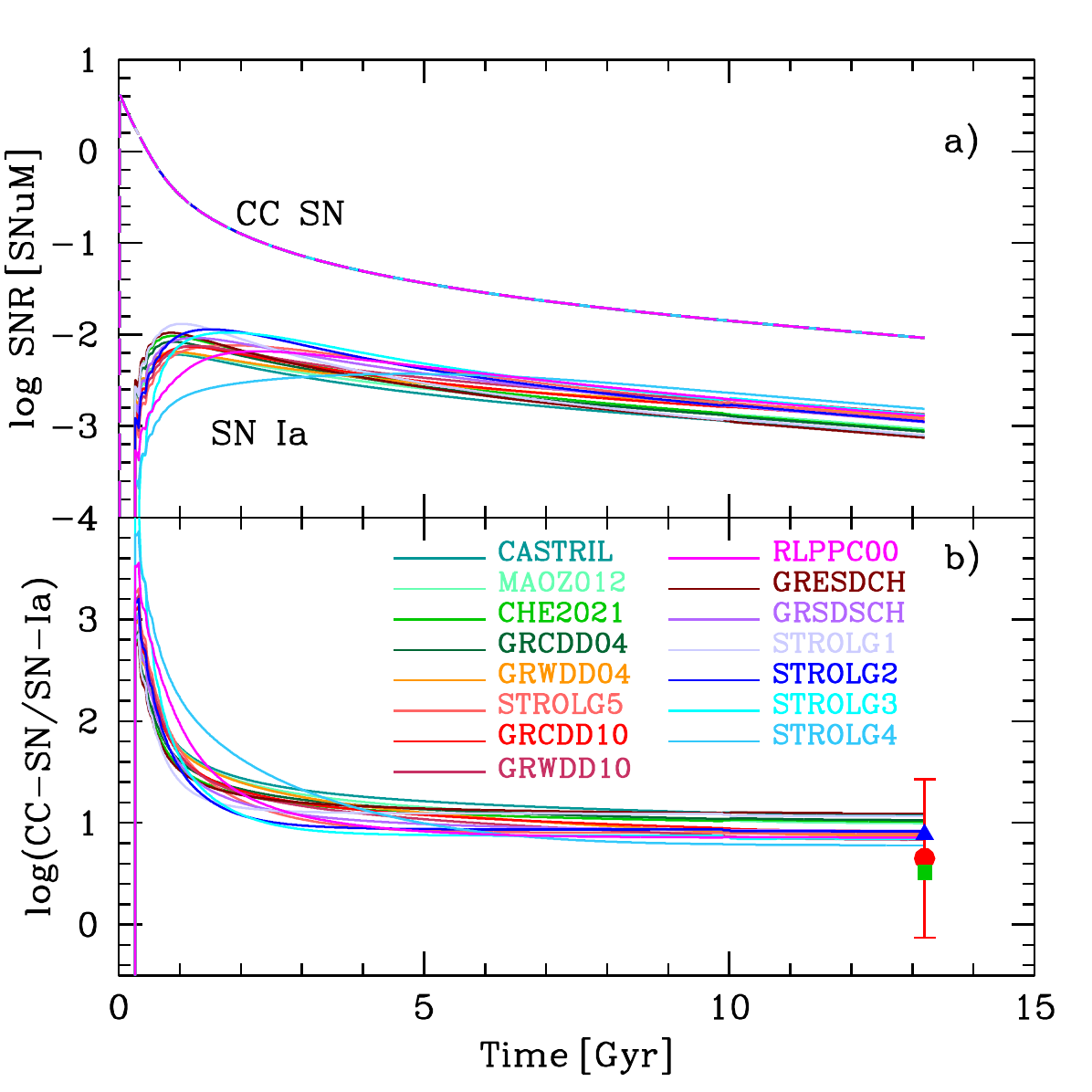}
\caption{
a) Time evolution of the SN rates, in SNuM units, for CC SN and SN Ia. The CC SN rates evolution fall in the same line for all models with a value of $\sim -2$ for the present time, about one dex above the SN Ia rate. The SN Ia rates vary, as shown different colours, depending on the used DTD.
b) Ratio between both SN-II/SN Ia, compared to different theoretical estimates  (blue and green dots from \citet{M17,Greggio2005}, respectively) and observational estimate (red dot with error bar) calculated as the mean and the standard deviation of values from \citet{Mannucci2005,Maoz2014,Li2011a}.}

\label{snt}
\end{figure}

Once our models are well calibrated, we proceed to analyse the results related to the SN rates. Panel a) of Fig.~\ref{snt} shows the evolution of both the CC SN and SN Ia rate as a function of time.
Since the same IMF is used for all models, the CC SN rate is basically the same for all of them, as shown by the (overlapping) lines located in the upper side of the panel a).
In contrast, the SN Ia rate evolves differently in each model, according to the used DTD. Thus, SN Ia appear at a different time for each DTD 
(see delay times in Table~\ref{tab:dtdmodels}), although it is difficult to see it at the scale of the age of the Universe. The first to appear is the STROLG1 (lavender line), followed by GRSDSCH, CASTRIL, and MAOZ017, while the last ones are RLPPCC00, STROLG3 and STROLG4.  In panel b) it is shown how the ratio between both types of SN changes among models. The lowest value at the present time corresponds to STROLG4 (0.78) and the highest one is for GRSDSCH (1.09). In the same panel, we also represent the ratio $N_{\rm CC}/N_{\rm Ia}$ from \citet{M17}, as a green square, and the ratio around $\sim 5\text{---}7$ (depending on the IMF) advocated by \citet{Greggio2005} for a SSP as a blue triangle. 

From an observational point of view, this ratio is dependent on the selected galaxy sample, varying with morphological type and $B-V$ colour \citep{Li2011b}. Being affected by the star formation history, we do find that it changes according to the radial region, although we only represent here the solar region. Following \citet{Mannucci2005,Maoz2014,Li2011a}, the observed ratio between CC and Ia SNe may be as small as 1 (or even lower than 1) or as large as $\sim 200$. This range is illustrated by a red dot with error bars. Thus, all used DTDs give equally acceptable results for the present time within the range defined by the error bars.

\begin{figure}
\centering
\includegraphics[width=0.47\textwidth]{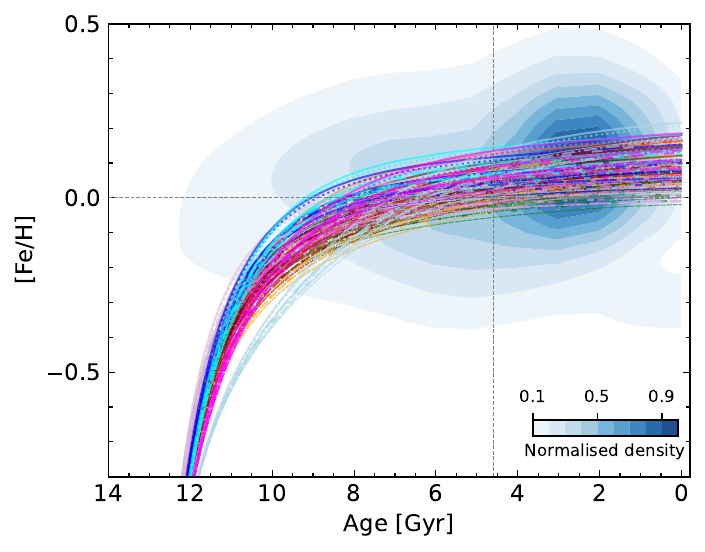}
\caption{The Age evolution of metallicity, as Iron abundance [Fe/H] for the solar vicinity. The observational data as given by Fig.~\ref{feht_obs} and Section \ref{obs} are represented by the contour plots for a better visualisation. The dotted lines indicate the Sun position at its birth time, 4.6\,Gyr ago, for reference. Different models are plotted as lines following the style and colour scheme given in Tables \ref{tab:dtdmodels} and \ref{tab:sniayields} and Fig.~\ref{snt}. Only the 148 {\sl realistic} models are drawn.
}
\label{feht}
\end{figure}

\subsection{The time evolution of metallicity and alpha-over-iron relative abundance on the solar vicinity}
\label{sec:feh_xfe}

In this subsection the results related with the time evolution of the abundance for Iron [Fe/H] and 
$\alpha$-elements [X/Fe] for the 148 models considered as {\sl realistic} are presented. In Fig.~\ref{feht}, we display the age-metallicity relation for the solar region, the metallicity being represented by [Fe/H]. Some combinations reproduce the data better than others, however, it is very difficult to choose the best model, given the high dispersion of data.

\begin{figure*}
\centering
\includegraphics[width=0.9\textwidth]{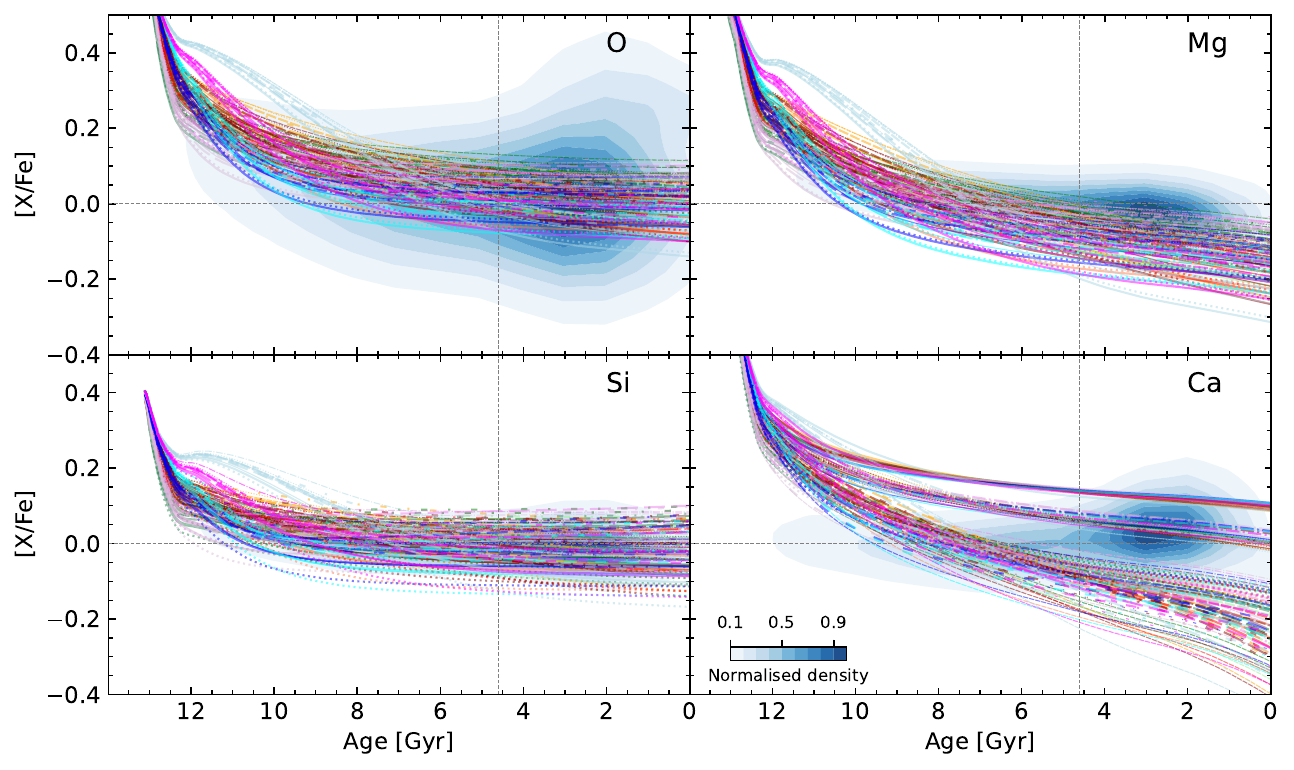}
\caption{The relative abundances of [O/Fe], [Mg/Fe], [Si/Fe], and [Ca/Fe] as a function of the stellar age in Gyr. Different models are represented by different colours and types of lines following the same style and colour scheme as in Fig.~\ref{feht}. The observational data as given by Fig.~\ref{xfe_age_obs} and Section \ref{obs} are represent by the contour plots for a better visualisation. The dashed lines indicate the Sun position for reference.
}
\label{fig:xfe_age_148_models}
\end{figure*}

\begin{figure*}
\centering
\includegraphics[width=0.9\textwidth]{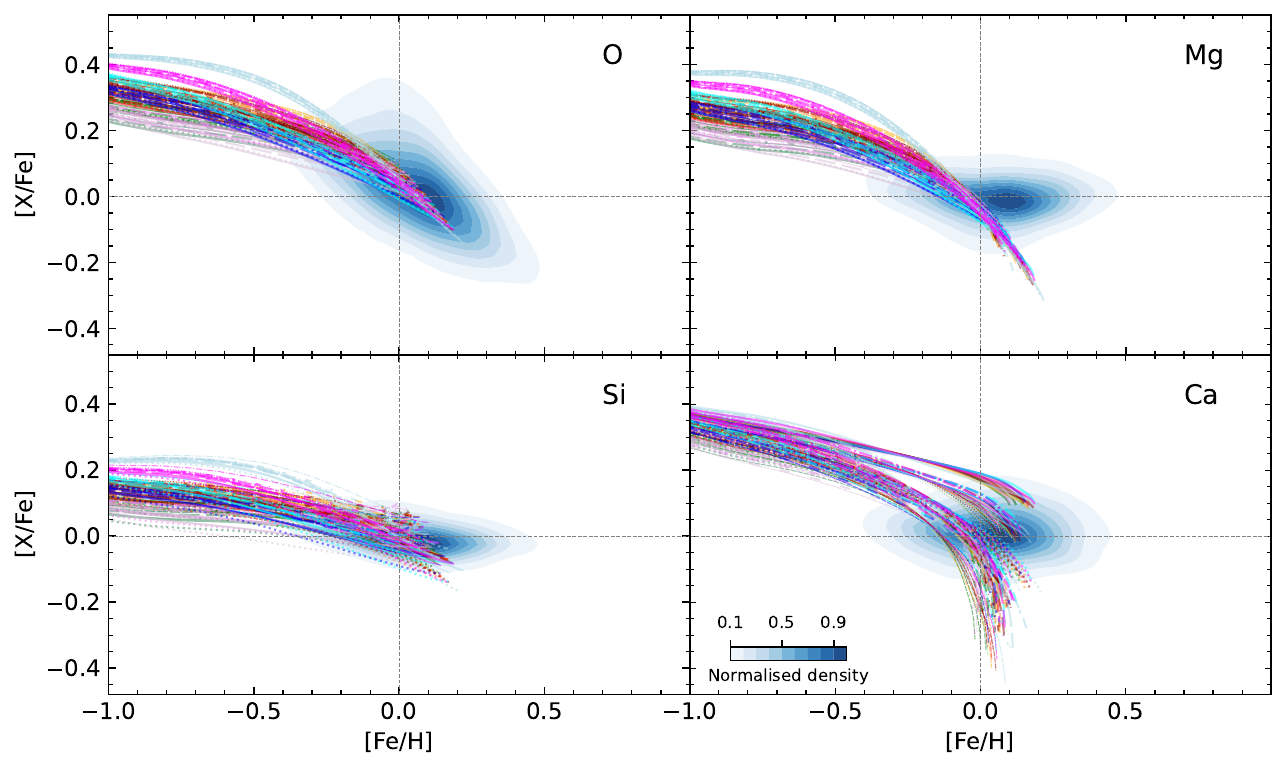}
\caption{The same as Fig. \ref{fig:xfe_age_148_models} for the relative abundances of [O/Fe], [Mg/Fe], [Si/Fe], and [Ca/Fe] as a function of [Fe/H].}
\label{fig:xfe_fe_148_models}
\end{figure*}

The next step is to analyse the evolution of different 
$\alpha$-elements relative abundances [X/Fe], which incorporate both CC and Ia SNe ejecta. 
In Fig.~\ref{fig:xfe_age_148_models} we present the relative abundances [X/Fe] for elements O, Mg, Si, and Ca as a function of the stellar age in Gyr. In all plots, each set of a different colours shows models for a given DTD (with the same colour meaning as in previous figures). As before, the different yields from SN Ia have a smaller effect over the plots compared with the DTD and the results show a dispersion around each colour/DTD. From the solar age (dashed vertical line) to the present age, all models follow a similar behaviour with differences only in the absolute abundances for O, Mg and Si, while Ca shows quite important differences in this last evolution times for some particular SN Ia yield sets. For Ca, LEU2020 gives results well down of the data points from the solar birth time until the present time for all DTDs.

Since the elements produced by massive stars appear first in the ISM, the relative abundances [X/Fe] are high compared to the solar values for early times/old stellar ages. Later, the abundances of the iron-peak elements, produced by SNe Ia, start to increase in the ISM and the relative abundances decrease. From an evolutionary point of view the relative abundances [X/Fe] begin to decrease just when the first SN Ia explode, and, obviously the time when this occurs is very dependent on the DTD. Following these plots, models using CASTRIL and STROLG1 are the closest ones to the data for ages larger than 10\,Gyr (early times). This is expected since these DTDs produce SN Ia very promptly once binary systems are created, as indicated by the short delay times $\Delta t = 40$\,Myr in the first case, and by the parameter  $\upxi=10$\,Myr in the second one. The STROLG4 models --light blue lines-- DTD show a bump for all elements at early times (old ages) that does not exist in the data trends doing this DTD invalid for our purposes.

Fig.~\ref{fig:xfe_fe_148_models} displays the same relative abundances [X/Fe] but now as a function of [Fe/H], following the same structure as before. To use [Fe/H] as a proxy of time has the advantage of eliminating the stellar age uncertainties, since only abundances are involved. \citet[ see their Figs. 17 and 18]{kobayashi2020}, considering a MCh case and a sub-MCh case -- associated to SD and DD scenarios, respectively--, claim that in the plot of [O/Fe] {\it vs.} [Fe/H] the difference between SD-DTDs and DD-DTDs would appear as a flatter line in the lower metallicities with a strong decline for around solar metallicities in the first case, while for the second case the line would decline smoothly from the lowest metallicities. We see, however, that  not only these two behaviours appear for our set of models with different DTDs, and different slopes are observed. The flatter line is the one from STROLG4 --light blue lines--, while the steeper ones correspond to STROLG1, STROLG2 and STROLG3, --lavender, blue and cyan lines--, all of them corresponding to a SD mechanism. The plateau also appears for STROLG5, GRWDD04 and RLPPC00 with a DD scenario.  These results are in agreement with the ones obtained by \citet{Vincenzo2017} and, more recently, by \citet{Palicio2023}. Moreover, \citet{kobayashi2020}, by computing models with some different proportions of both scenarios, obtained that a combination with a $25\%$ sub-MCh SNe Ia and a $75\%$ MCh SNe Ia reproduces better the observational constraints that only one scenario.

Not all models are broadly compatible with the observational data. In fact, it seems difficult to find -at naked eye-- a model able to reproduce simultaneously the data in all panels (see next sections).  Thus, the best models for Ca seem to be the ones in the lower part of plots, (STROLG1 models, lavender lines), while O, Mg and Si data are better fitted with STROLG5 (coral lines), and other DTDs corresponding to DD scenarios as GRWDD04 (orange lines) , or RLPPC00 models (magenta lines), but also by STROLG2 and STROLG3 (blue and cyan lines), or GRESDCH (violet lines), which corresponds to a SD scenario. For Ca, the two sets of yields that in Fig. \ref{fig:xfe_age_148_models} show values above and below the observational data at lower ages, also show in Fig. \ref{fig:xfe_fe_148_models} the same disagreement with observations.

We see in Figs.~\ref{feht} and \ref{fig:xfe_fe_148_models} that the models do not reach [Fe/H] > 0.2\,dex as observed in the stellar data. Since the models show the enrichment only for the the solar region, there are two possibilities to explain stars with metallicity higher than 0.2\,dex: (i) they were born at other radial regions and then migrate to the solar region; (ii) the errors in the metallicities and the stellar ages for these stars are higher than the estimated ones. A more detailed investigation will be given in Section \ref{sec:disc} to support the first possibility.
%%%%%%%%%%%%%%%%%%%%%%%%%%%%%%%%%%%%%%%%%%%%%%%%%%%%%%%%%%%%%%%%%%%%%%%%%%%%%%%%%%%%%%%%%%%%%%%%%%%%%%%%%%%%%%%%%%%%
\subsection{Best models selection \label{sec:best_models}}

To select the best model capable of reproducing all our data simultaneously, we performed a maximum likelihood estimation \citep[see for example][]{LehmCase98}. We use the data described in Section \ref{obs} and compare them with the corresponding results in each model. To perform the calculation, we need to define the probability of an observation given a model. %Since the models are tabulated values for given input parameters, 
We first find the closest model point to the observational point of interest. The distance is taken in the space of all measurements available for a given observation, in our case the abundances and the age of each star.  Assuming that we have uncorrelated observations with normally distributed errors in the measurements, the probability of an observation for a given model is then given by:

\begin{equation}
\label{eq:obs-prob}
{P_{i}}(X_{obs} \,|\, M_{nmod}) = \max_{\forall m \in M_{nmod}} \prod_{j=1}^{N} \frac{1}{\sqrt{2\pi}\sigma_{i,j}} e^{-\frac{(X_{m,j}-X_{i,j})^{2}}{2\sigma_{i,j}^2}}, 
\end{equation}
where $M_{nmod}$ denotes a given model in the range $nmod \in [1,148]$, 
 $m$ is a model point in $M_{nmod}$, $X_{obs}$ represents each observational point $i$ that has $j$ measurements (abundances and age) with $\sigma_{i,j}$ their respective errors. %This approach is justified since we use the technique based on the Q-matrix formalism to compute the proportion of element ejected when the star dies. %As the abundances of the $\alpha$-elements are not correlated within the matrix, it is not expected any relation between [X/Fe] ratios.

We calculate for each defined model $M_{nmod}$ the likelihood using the following expression:

\begin{equation}
\label{eq:likelihood}
\mathcal{L}(M_{nmod} \,|\, X_{obs}) = \prod_{i=1}^{n}  P_{i}.
\end{equation}

\begin{figure}
\includegraphics[width=\columnwidth]{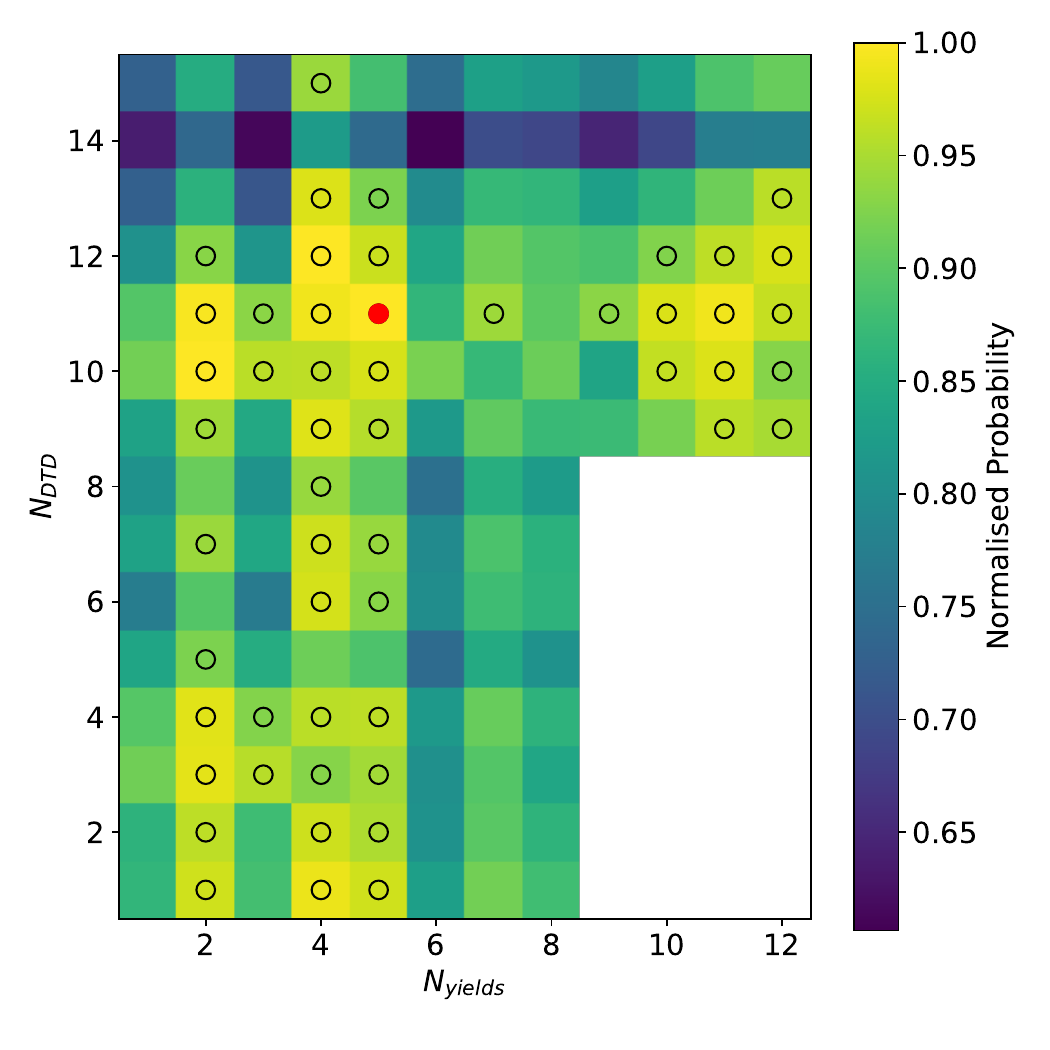}
\caption{The normalised probability distribution obtained based on the likelihood of each model in the range $nmod \in [1,148]$. Black empty circles highlight the models that were determined to be inside the  1$\sigma$ confidence level. The best model is marked by the red solid circle.
}
\label{fig:prob-distribution}
\end{figure}

Numerically, it is convenient to work with the log-likelihood and that is the procedure we adopt. With the calculated likelihood for each model on the grid, we can obtain the one that best explains the available data by finding the model that gives the maximum value. To obtain the models inside the 1$\sigma$ confidence level, as we are assuming a flat prior for all of them, we added the values of the highest-likelihood models that correspond to 39.3\% of the volume resulting from the cumulative distribution function. These results are summarised in Fig.~\ref{fig:prob-distribution}, where we show the probability distribution obtained based on the likelihood of each model in the range $nmod \in [1,148]$, with the models inside the $1\sigma$ confidence interval marked by empty black circles and the best model location marked by the red solid circle. Our analysis has shown that only at the 1$\sigma$ confidence level are there relevant differences between the models, such that this confidence interval will be considered in subsequent analyses.

\begin{figure}
\centering
\includegraphics[width=\columnwidth]{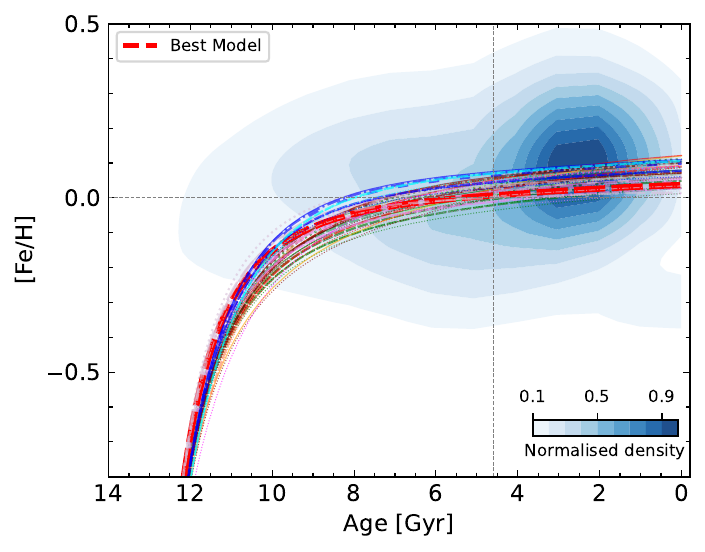}
\caption{[Fe/H] as a function of Age for the 52 best models selected from the statistical analysis at the 1$\sigma$ confidence interval. Different models are represented by different colours and types of lines following the same style and colour scheme as in Fig.~\ref{feht}. The dashed red line indicate the highest probability model as the best model.
}
\label{fig:fe_age_best_models}
\end{figure}
 
\begin{figure*}
\centering
\includegraphics[width=0.9\textwidth]{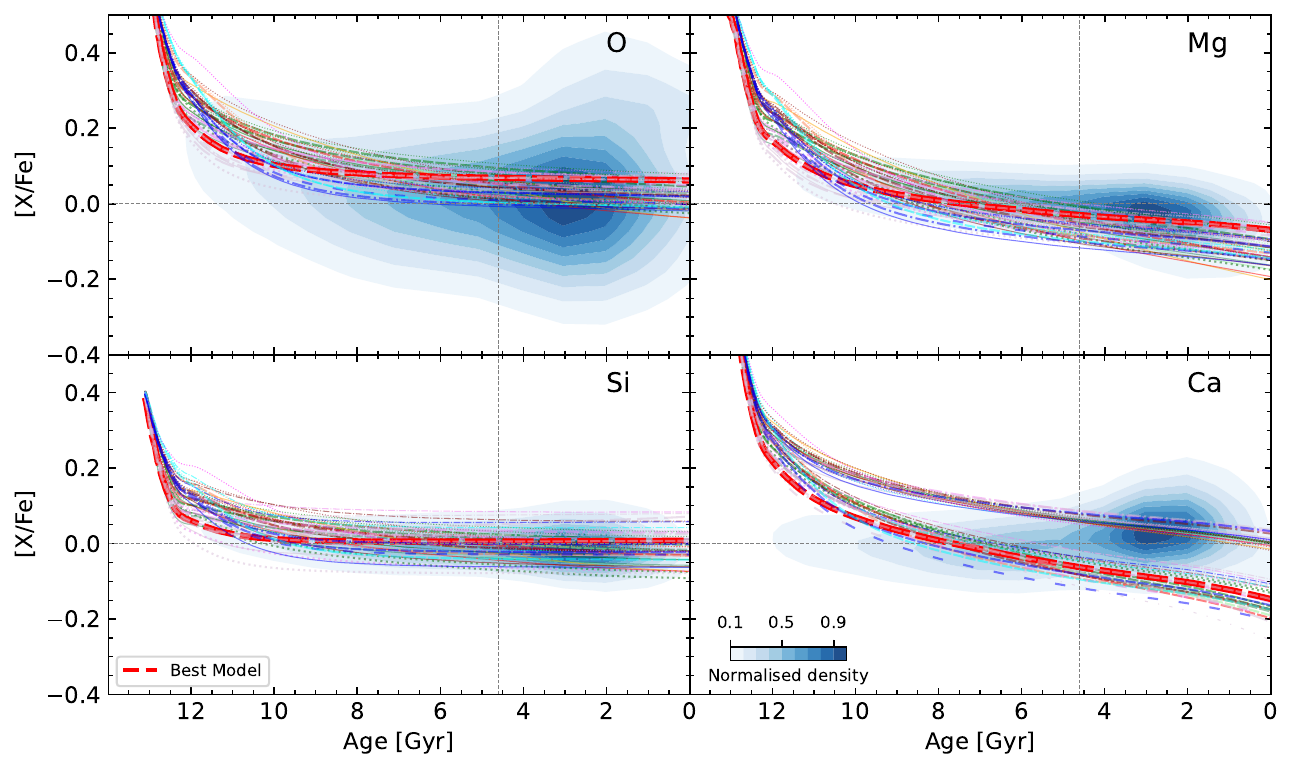}
\caption{As Fig. \ref{fig:xfe_age_148_models} but for only the 52 best models selected from the statistical analysis at the 1$\sigma$ confidence level. Different models are represented by different colours and types of lines following the same style and colour scheme as in Fig.~\ref{feht}. The dashed red line indicate the highest probability model as the best model.
}
\label{fig:xfe_age_best_models}
\end{figure*}

\begin{figure*}
\centering
\includegraphics[width=0.9\textwidth]{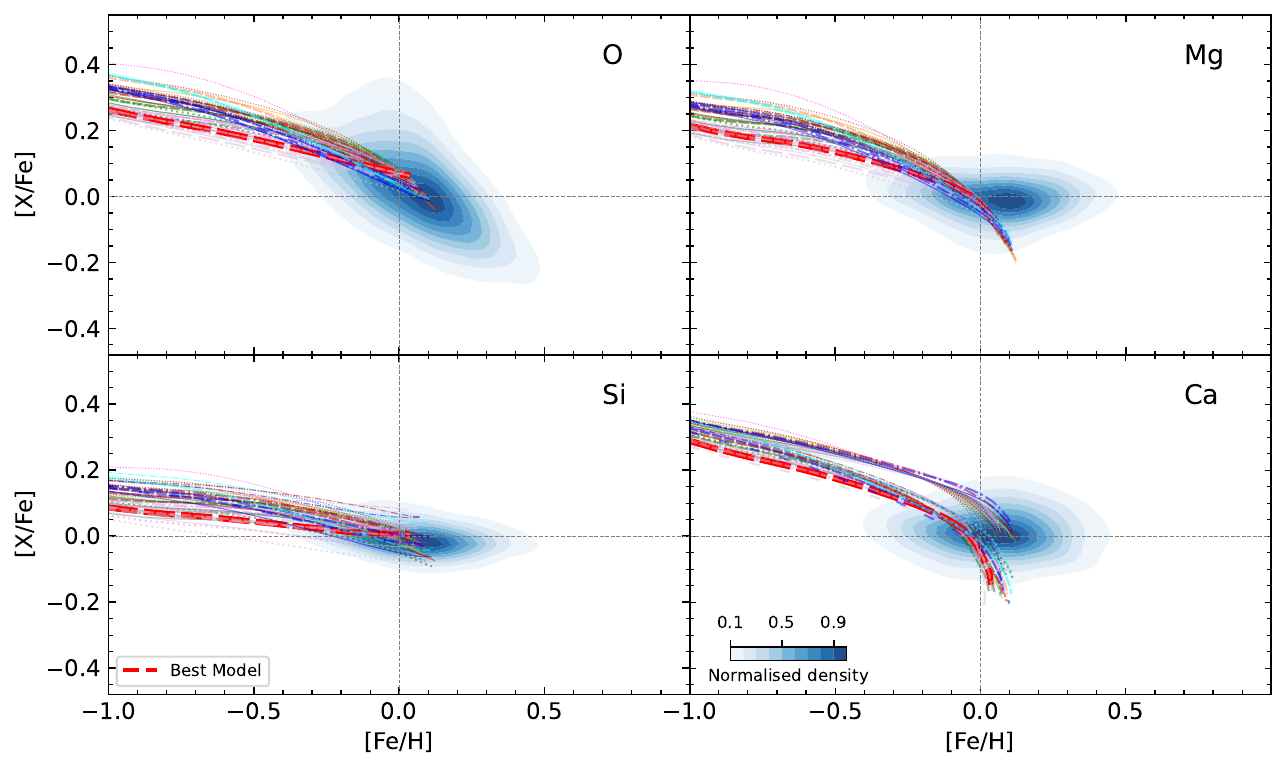}
\caption{As Fig. \ref{fig:xfe_age_best_models} but for [X/Fe] as a function of [Fe/H].
}
\label{fig:xfe_fe_best_models}
\end{figure*}

In Table~\ref{table:best_models} we give the best 52 models that at 1$\sigma$ confidence level are able to reproduce simultaneously the observational data. In this table, column 1 displays the DTD name, column 2 the type of scenario, column 3 the SN-Ia yields set and column 4 the explosion type. Column 5 displays the corresponding normalised probability associated with the maximum likelihood procedure. We include in Table~\ref{table:best_models}, column 6, the flag Q that gives the quality of the different Yield$+$DTD scenario compatibility, with six categories:

\begin{enumerate}[A)]
    \item The DTD and the yield are in good agreement from a theoretical perspective, corresponding to the classic SD-MCh-DDT scenario; 
    \item  These model combinations are also in good agreement from the theoretical perspective, although they correspond to the SD-sub-Mch-DDet scenario;
    \item Models where the yield was calculated by a parametrization associated to the SD-DDT scenario, like high central densities for the WD, while the DTD model came from a DD scenario;
    \item This case refers to models whose DTD were derived from observational analysis. For these models the yields correspond to different calculations of the same classic SD-MCh-DDT scenario, which may contradict the DD time dependence ($t^{-1}$) of the DTD models;
    \item Models whose yields are for a double-detonation mechanism at sub-MCh, but their associated DTD do not include the sub-MCh case as a possible SNIa progenitor. Although the yield and the DTD are both consistent with an SD scenario, in some combinations, the DTD models assume a hydrogen donor while the yields needs a helium donor star;
    \item Models where the yield corresponds to the classic Mch-DDT case, but the DTD assumes the SD scenario with a sub-Mch explosion mechanism.
\end{enumerate}

Categories A and B do not have any incompatibility between the yield and DTD, while the remaining present some warnings, as described. It is interesting to point out that the four best models in Table \ref{table:best_models} correspond to the combination DTD function and SN Ia yield that is in agreement with the classic SD and Mch scenario. Within the 52 models which may reproduce all the data plots simultaneously at the 1$\sigma$ confidence level, the results point to a higher predominance of lower N$_{yields, SN-Ia}$, indicating that models with yields that include MCh or Near MCh (classical scenario) correspond to 75\% and are most probable to fit the data. Analysing the DTDs distribution within the 52 most probable models, we see that 40\% are in the DD scenario, while the remaining 60\% correspond to the SD scenario. We also noticed that in Fig. \ref{fig:prob-distribution} there are three regions that concentrate the highest probabilities, approximately located in the intervals: 

\begin{enumerate}[1)]
    \item N$_{\rm yields, SN-Ia} \in [2,5]$ and N$_{\rm DTDs} \in [2,4]$;
    \item N$_{\rm yields, SN-Ia} \in [2,5]$ and N$_{\rm DTDs} \in [9,13]$; 
    \item N$_{\rm yields, SN-Ia} \in [10,12]$ and N$_{\rm DTDs} \in [10,12]$.
\end{enumerate}

The results of our multidimensional analysis independently confirm previous results in the literature, as those from \citet{kobayashi2020}, where by the analysis of [O/Fe] {\it vs.} [Fe/H] and considering only the MCh case and a sub-MCh cases, it is found that a combination with a $25\%$ sub-MCh SNe Ia and a $75\%$ MCh SNe Ia reasonably reproduces the observational constraints.

At the 1$\sigma$ confidence level, we can discard the SN Ia yields BR20191, LN20182 and IW1999 (MCh explosions) and the DTD function STROLG4 (SD), that for the whole set of combinations of SN Ia yields and DTDs functions proposed in this work, none has satisfactory results for them. 

The results of the 52 best models are displayed in Figs. \ref{fig:fe_age_best_models}, \ref{fig:xfe_age_best_models}
and \ref{fig:xfe_fe_best_models}, for [Fe/H] {\it vs.} Age, [X/Fe] {\it vs.} Age and [X/Fe] {\it vs.} [Fe/H], respectively. We also plotted in these figures the best model selected as the highest probability model, given by the combination N$_{yields, SN-Ia}= 5$ (LN20181, Near MCh scenario) and N$_{DTD}=11$ (STROLG1, SD scenario). Considering the age evolution (Figs. \ref{fig:fe_age_best_models} and \ref{fig:xfe_age_best_models}), the best models are a compromise between fitting the data at large ages values and also the high density data at $\sim 3$\,Gyr. For Ca, a fraction of the best models fits the higher ages data and the other fraction the high density data at $\sim 3$\,Gyr. Considering the abundances ratios as a function of [Fe/H] (Fig. \ref{fig:xfe_fe_best_models}), most of the best models pass through the high density data at [Fe/H] $\sim 0.0$ dex, with the exception of some models for Mg and Ca. By analysing these figures, some part of the data dispersion in the [X/Fe] ratios may be explained by the mixing of different SN Ia production channels and yields. This will be addressed in the Section \ref{sec:disc}.

Ultimately, the combination of SN Ia yields and DTD functions in Table \ref{table:best_models} can be used as reference for future chemical evolution models that intend to reproduce the chemical abundances in the Galactic disc. 

\begin{table}
\setlength{\tabcolsep}{4pt}
\centering
\caption{The best 52 models selected from
the statistical analysis at the 1$\sigma$ confidence level.}
\label{table:best_models}
%\resizebox{15cm}{!}{
\begin{tabular}{lccccc}
\hline											
DTD	&	Scenario	& 	Yields-SN-Ia	&	Expl. type	&	N. Prob.	&	Q	\\
(1)	&	(2)	&	(3)	&	(4)	&	(5)	&	(6)	\\
\hline											
STROLG1 & SD & LN20181 & Mch & 1.00000 & A \\
STROLG2 & SD & MO20182 & Mch & 0.99997 & A \\
GRESDCH & SD & SEI2013 & Mch & 0.99911 & A \\
STROLG1 & SD & SEI2013 & Mch & 0.99652 & A \\
STROLG1 & SD & GR20211 & sub-Mch & 0.99227 & E \\
STROLG1 & SD & MO20182 & Mch & 0.99200 & A \\
CASTRIL & DD & MO20182 & Mch & 0.98826 & D \\
CHE2021 & DD & SEI2013 & Mch & 0.98453 & D \\
GRCDD04 & DD & SEI2013 & Mch & 0.98297 & C \\
GRSDSCH & SD & MO20182 & Mch & 0.98127 & F \\
GRESDCH & SD & GR20211 & sub-Mch & 0.97925 & E \\
STROLG3 & SD & MO20182 & Mch & 0.97914 & A \\
STROLG1 & SD & LEU2020 & sub-Mch & 0.97847 & E \\
GRESDCH & SD & LN20181 & Mch & 0.97626 & A \\
STROLG2 & SD & GR20212 & sub-Mch & 0.97573 & E \\
STROLG5 & DD & MO20182 & Mch & 0.97432 & C \\
CASTRIL & DD & LN20181 & Mch & 0.97186 & D \\
CASTRIL & DD & SEI2013 & Mch & 0.97138 & D \\
GRCDD10 & DD & MO20182 & Mch & 0.97079 & C \\
MAOZ017 & DD & MO20182 & Mch & 0.97053 & D \\
STROLG2 & SD & LN20181 & Mch & 0.96914 & A \\
STROLG1 & SD & GR20212 & sub-Mch & 0.96605 & E \\
GRESDCH & SD & LEU2020 & sub-Mch & 0.96422 & E \\
MAOZ017 & DD & SEI2013 & Mch & 0.96146 & D \\
GRCDD04 & DD & LN20181 & Mch & 0.96044 & C \\
STROLG2 & SD & GR20211 & sub-Mch & 0.96017 & E \\
GRESDCH & SD & MO20182 & Mch & 0.96010 & A \\
GRSDSCH & SD & GR20211 & sub-Mch & 0.95940 & B \\
STROLG3 & SD & GR20212 & sub-Mch & 0.95918 & E \\
GRCDD04 & DD & MO20182 & Mch & 0.95906 & C \\
GRESDCH & SD & MO20181 & Mch & 0.95881 & A \\
CHE2021 & DD & MO20181 & Mch & 0.95842 & D \\
GRSDSCH & SD & LN20181 & Mch & 0.95642 & F \\
MAOZ017 & DD & LN20181 & Mch & 0.95214 & D \\
GRSDSCH & SD & GR20212 & sub-Mch & 0.94882 & B \\
CHE2021 & DD & LN20181 & Mch & 0.94489 & D \\
GRSDSCH & SD & SEI2013 & Mch & 0.94461 & F \\
STROLG1 & SD & LN20183 & Mch & 0.94212 & A \\
RLPPC00 & Mix & MO20182 & Mch & 0.94116 & A \\
GRCDD10 & DD & SEI2013 & Mch & 0.94032 & C \\
GRWDD10 & DD & MO20182 & Mch & 0.93988 & C \\
GRCDD10 & DD & LN20181 & Mch & 0.93929 & C \\
STROLG1 & SD & BR20192 & sub-Mch & 0.93209 & E \\
STROLG1 & SD & MO20181 & Mch & 0.93083 & A \\
STROLG5 & DD & LN20181 & Mch & 0.93039 & C \\
STROLG2 & SD & SEI2013 & Mch & 0.93031 & A \\
GRESDCH & SD & GR20212 & sub-Mch & 0.92906 & E \\
CHE2021 & DD & MO20182 & Mch & 0.92797 & D \\
GRCDD04 & DD & MO20181 & Mch & 0.92767 & C \\
STROLG2 & SD & LEU2020 & sub-Mch & 0.92475 & E \\
STROLG3 & SD & LN20181 & Mch & 0.92203 & A \\
GRWDD04 & DD & SEI2013 & Mch & 0.92174 & C \\
\hline																					
\end{tabular}
%}
\end{table}

%%%%%%%%%%%%%%%%%%%%%%%%%%%%%%%%%%%%%%%%%%%%%%%%%%%%%%%%%%%%%%%%%%%%%%%%%%%%%%%%%%%%%%%%%%%%%%%%%%%%%%%%%%%%%%%%%%%%

\subsection{Results for other radial regions and abundances dispersions}
\label{sec:disc}

As obtained in Section \ref{sec:best_models}, 52 best models are able to reproduce all observational data sets simultaneously, considering the 1$\sigma$ confidence level. It is, therefore, possible that some channels (DTDs or explosion mechanisms) appear simultaneously considering all SN Ia events in a given region. In that case, they would also create some data dispersion. As we noted previously, the relative abundance [X/Fe] data for the $\alpha$-elements show a large dispersion that any chemical evolution model may not reproduce with only one region. In order to seek for a possible cause of the high dispersion in the observational data, we look at the dispersion that the 52 best models produce when other radial regions are included in the plot.

\begin{figure}
\centering
\includegraphics[width=\columnwidth]{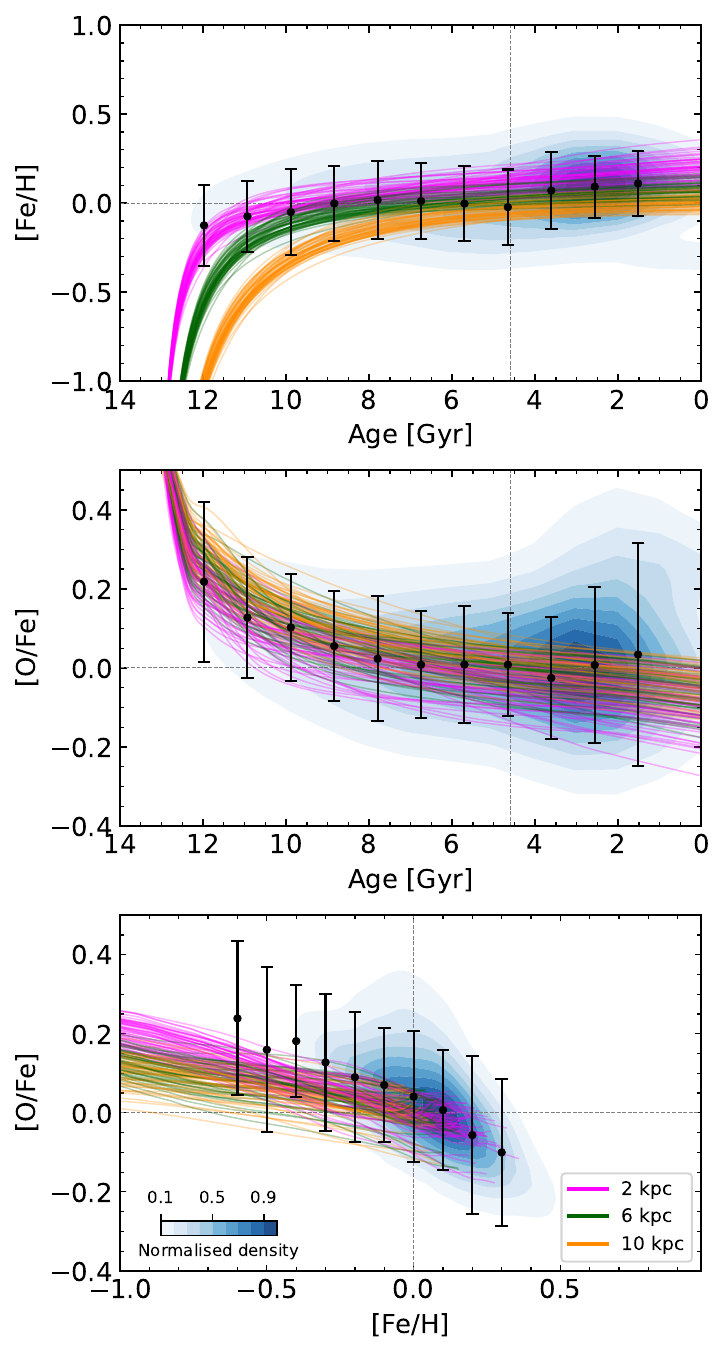}
\caption{Results for the 52 best models and considering the radial regions 2, 6 and 10\,kpc for each model for: [Fe/H] {\sl vs.} Age (top panel), [O/Fe] {\sl vs.} Age (middle panel) and [O/Fe] {\sl vs.} [Fe/H] (bottom panel). Colour lines for the models represent different radial regions as labelled in the bottom panel. The contour plots represent the data described in Section \ref{obs} and in the previous figures. The binned data are shown by the black dots with error bars and correspond to the mean and the standard deviation of the data, respectively, weighted by the respective errors.}
\label{fig:results_all_radii}
\end{figure}

The age-metallicity relation is represented in Fig.~\ref{fig:results_all_radii} top panel. There is a clear dependence on the Galactocentric distance, the inner regions showing higher metallicities than the outer ones for the same age. This results from the negative radial gradient of metallicity presented at the Galactic disc and this dependency is stronger in the past than now. In Fig.~\ref{fig:results_all_radii} middle panel, we represent the evolution of [O/Fe] as a function of the stellar age for the 52 best models and including different radial regions of the disc, as in the top panel. In the same figure, the bottom panel shows the relation [O/Fe] {\sl vs.} [Fe/H]. The binned data are shown by the black dots with error
bars and correspond to the mean and the standard deviation of the data, respectively, weighted by the respective errors. The fact that different radial regions show variable [Fe/H] and [O/Fe] is due to the SFR changing on radius in the disc, with higher rates in the early times for the inner regions, while in the outer regions the SFR starts with low values increasing with time.

Since different regions of the disc do show different patterns in the plane [Fe/H] {\sl  vs.} Age and in the plane [X/Fe], either {\sl  vs.} Age or {\sl vs.} [Fe/H], the observed dispersion would be an indicator of mixing of stars originally born in different radial regions than the solar neighbourhood. Additionally, the results point to an inner radial migration towards the solar region. The different DTDs or SN-Ia yields sets may also contribute to the dispersion. These results agree with  \citet{palla2022} findings, who by using [Mg/Fe] abundances from AMBRE-HARPS, and comparing them with detailed chemical evolution models, conclude that none set of yields may reproduce the data, even taken into account the stellar migration. We note that \citet{Buder2019} also find evidence for radial migration in the GALAH data at the solar vicinity, where stars older than $8$~Gyr have angular momenta lower than the Sun, which implies that they are in eccentric orbits and are originated in the inner disc.  

\section{Conclusions}
\label{Conclusions}

In this paper we have examined the role of Type Ia supernovae, focusing primarily on the different Delay Time Distributions (DTDs) determined by the scenarios of binary systems, as well as the yields of elements created by different explosion mechanisms. The findings of this study can be summarized as follows:

\begin{itemize}
\item We have analysed carefully a large set of observations related with the $\alpha$-element over Iron abundances obtained from some modern surveys and other data of the literature. The detailed description of these data can be seen at Appendix~\ref{AppA}.

\item An updated version of the {\sc MulChem} chemical evolution model applied to the Milky Way Galaxy was used, where we have now taken the stellar yields from \citet{cris11,cris15} (the FRUITY set), for low and intermediate mass stars, and from \citet{lim18} for massive stars with several rotation stellar velocities. The calibration of the MWG model by using these stellar yields is given in Appendix~\ref{AppB}.

\item To these prescriptions, we have added 15 possible DTDs for both  DD and SD scenarios, and 12 sets of SNe Ia yields from recent works with different mechanisms of explosion. The combination of both inputs provide 180 different models, 148 of them considered {\sl realistic} enough, although some of them are better than others, considering the current knowledge of SNe Ia production. The models results have been compared with the observational data presented in the first part of work. 

\item For this comparison, a multidimensional maximum likelihood procedure was implemented, where the abundances of O, Mg, Si, Ca and Fe and, additionally, the stellar age are fitted simultaneously. From the analysis, 52 best models are able to reproduce the datasets simultaneously at the 1$\sigma$ confidence level, which represent a $\sim 35$\% of the realistic models (a $\sim 29\%$ of the total number of computed models). From these best models, the most probable DTD is the one called STROLG1 from \citet{strolger}, defined by the parameters $\upxi=10$\,Myr, $\omega=600$, $\alpha=220$, with a very prompt delay. The best set of SN Ia yields seems to be LN20181 (MCh) from \citet{leung2018}. The combination STROLG1 (SD) $+$ LN20181 could indicate that a single degenerate with a time of delay until the explosion from the merger (SD with MED) could be a real channel for the SN Ia production. 

\item Within the 52 best models able to reproduce all observational data sets simultaneously, considering the 1$\sigma$ confidence level, the SN Ia yields that include MCh or Near MCh (classical scenario) correspond to 75\% and are most probable to fit the data. For the DTDs within the 52 most probable models, 60\% correspond to the SD scenario, while the remaining 40\% are in the DD scenario. It is, therefore, possible that some of these channels (DTDs or explosion mechanisms), taking into consideration the 52 best models, appear simultaneously considering all SN Ia events in a given region. 

\item The dispersion given by chemical evolution models is high for different radial regions of the disc. This might suggest that the observed dispersion is related to uncertainties in distance determinations, or with the stellar migration in the disc. In our models, differences between radial regions are due to the SFR variations. The data dispersion may be also explained by a mix of DTDs (different scenarios to create SN Ia binary systems), or by a mix of explosions channels. The observed dispersion is probably due to both effects: radial mixing and several simultaneous SN Ia scenarios/channels.

\end{itemize}

\section*{Acknowledgements}

We are very grateful to the reviewer whose comments and suggestions have helped to improve the manuscript.

This work is result of the grants  PGC-2018- 0913741-B-C22, MDM-2015-0500, MDM-2017-0737 (Unidad de Excelencia Mar{\'\i}a de Maeztu CAB), and PID2019-107408GB-C41, 
PID2019-107408GB-C42, PID2022-136598NB-C33, funded by MCIN/AEI/10.13039/501100011033. O.C. acknowledges funding support from FAPEMIG grant APQ-00915-18.
L.G. acknowledges financial support from AGAUR, CSIC, MCIN and AEI 10.13039/501100011033 under projects PID2020-115253GA-I00, PIE 20215AT016, CEX2020-001058-M, and 2021-SGR-01270.

This work has made use of the computing facilities available at the Laboratory of Computational Astrophysics of the Universidade Federal de Itajubá (LAC-UNIFEI). The LAC-UNIFEI is maintained with grants from CAPES, CNPq and FAPEMIG.

This work made use of the Third Data Release of the GALAH Survey (Buder et al. 2021). The GALAH Survey is based on data acquired through the Australian Astronomical Observatory, under programs: A/2013B/13 (The GALAH pilot survey); A/2014A/25, A/2015A/19, A2017A/18 (The GALAH survey phase 1); A2018A/18 (Open clusters with HERMES); A2019A/1 (Hierarchical star formation in Ori OB1); A2019A/15 (The GALAH survey phase 2); A/2015B/19, A/2016A/22, A/2016B/10, A/2017B/16, A/2018B/15 (The HERMES-TESS program); and A/2015A/3, A/2015B/1, A/2015B/19, A/2016A/22, A/2016B/12, A/2017A/14 (The HERMES K2-follow-up program). We acknowledge the traditional owners of the land on which the AAT stands, the Gamilaraay people, and pay our respects to elders past and present. This paper includes data that has been provided by AAO Data Central (datacentral.org.au).

This work has made use of data from the European Space Agency (ESA) mission
{\it Gaia} (\url{https://www.cosmos.esa.int/gaia}), processed by the {\it Gaia}
Data Processing and Analysis Consortium (DPAC,
\url{https://www.cosmos.esa.int/web/gaia/dpac/consortium}). Funding for the DPAC
has been provided by national institutions, in particular the institutions
participating in the {\it Gaia} Multilateral Agreement.

\section{Data availability}
\begin{enumerate}
    \item[1] The Appendix~\ref{AppA}, Supporting Information, gives the whole description of the data and the normalisation of different data sets.
    \item[2] The whole set of 180 models are available in electronic format upon request to the authors. 
    \item[3] The DTD functions and the SN Ia yields are included in the package {\sc starmatrix}, an Open Source Python code \citep{starmatrix} that compute the new ejected elements by each SSP, (see \url{https://github.com/xuanxu/starmatrix}).
    \item[4] The calibration of the Galactic chemical evolution model is given in Appendix~\ref{AppB}, Supporting Information.

\end{enumerate}

\bibliographystyle{mnras}

\bibliography{biblio}

%\clearpage
\appendix
\section{Normalised data}
\label{AppA}

The description of the datasets used in this work are provided in Table~\ref{table:obs}. The main point is that these abundances are given as [X/Y] $=\log{(X/Y)}-\log{(X/Y)_{\sun}}$, and obtained by using different solar abundances, as shown in Table~\ref{AppA:solar}. Therefore, we have re-normalised them  to the same solar abundances scale from \citet{Lodders2019}. 

\begin{table*}
\caption{Solar reference abundances.}
\label{AppA:solar}
\begin{tabular}{ccccccccccc}
\hline
Reference*	&	 Name 	&	Solar$\dagger$	&	$\epsilon({\rm Fe})_{\sun}$	&	$\epsilon({\rm O})_{\sun}$	&	$\epsilon({\rm Mg})_{\sun}$	&	$\epsilon({\rm Si})_{\sun}$	&	$\epsilon({\rm S})_{\sun}$	&	$\epsilon({\rm Ca})_{\sun}$	& $\epsilon({\rm C})_{\sun}$ & $\epsilon({\rm N})_{\sun}$ 	\\
\hline
(1)	&	CHEN00	&	A+09	&	7.50	&	8.69	&	7.60	&	7.51	&	--	&	6.34 & -- & --	\\
(2)	&	HARPS-GTO	&	B+15	&	7.47	&	8.71$^{\rm a}$	&	--	&	--	&	--	&	--	&	--	&	--	\\
(3)	&	HARPS-GTO	&	S+00, A\&G89	&	7.47$^{\rm b}$	&	--	&	--	&	--	&	--	&	--	&	--	&	8.05	\\
(4)	&	HARPS-GTO	&	C+11, B+15	&	7.52$^{\rm c}$	&	8.71	&	--	&	--	&	--	&	--	&	8.50	&	--	\\
(5,6)	&	HARPS-GTO	&	S+00, A\&G89	&	7.47$^{\rm b}$	&	8.93	&	7.58	&	7.55	&	--	&	6.36	&	--	&	--	\\
(7)	&	GALAH	&	G+07	&	7.45	&	8.56	&	7.53	&	7.51	&	--	&	6.31	&	8.39	&	--	\\
--	&	--	&	L+19	& 	7.54	& 	8.82	& 	7.61	&  	7.60	& 	7.24	& 	6.36	&	8.56	&	7.94	\\
\hline
\end{tabular}
\begin{minipage}{15cm}
\flushleft{
\footnotesize{* As in Table \ref{table:obs}. \\
$\dagger$ A\&G89: \citet{anders89}; A+09:\citet{asplund09}; G+07: \citet{grevesse07}; B+15: \citet{bertran2015}; S+00: \citet{santos00}; C+11: \citet{caffau11}; L+19: \citet{Lodders2019}.\\
$^{\rm a}$ OI 6158.171.\\
$^{\rm b}$ From S+00.\\
$^{\rm c}$ From C+11. }
}
\end{minipage}
\end{table*}

After this, we have still seen differences in their scale, since we consider that all abundances [X/Fe] must be zero when the Iron abundance or metallicity [Fe/H]=0 and when the stellar age $\tau= 4.6$\,Gyr, that is, the data must be around the solar values in the solar metallicity and age. We have, therefore, analysed each cloud of data for each set and element, to search for its center and then to shift it to locate it in the solar value.

\begin{figure*}
\includegraphics[width=0.55\textwidth,angle=0]{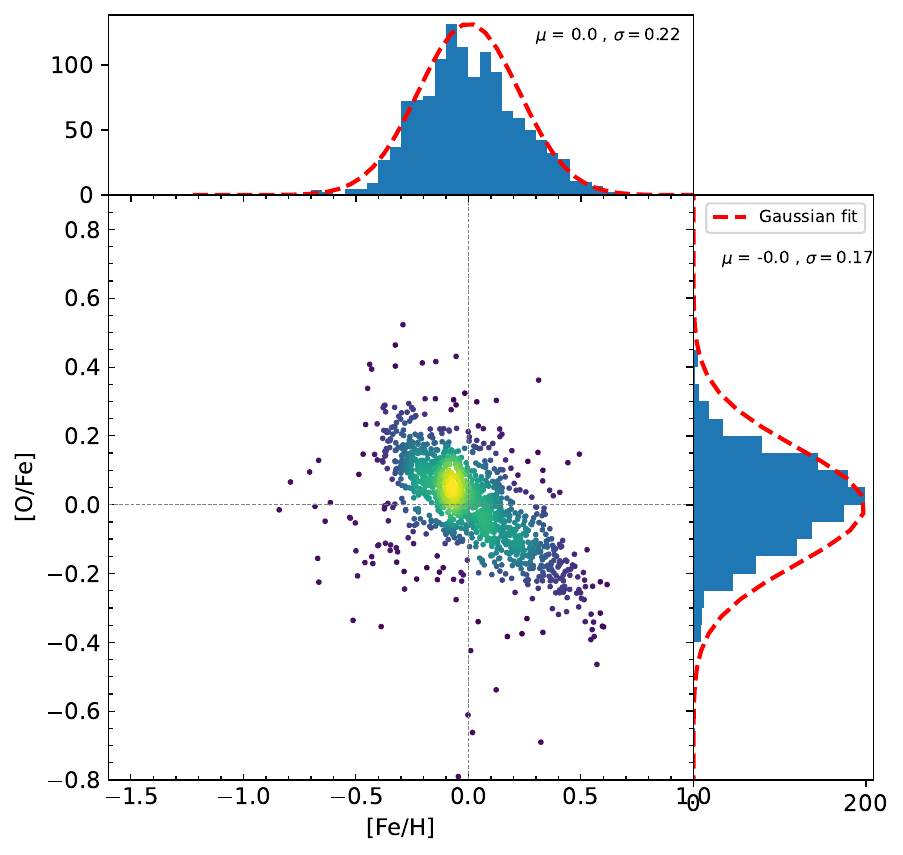}
\caption{Distribution of the abundances [O/Fe] vs. [Fe/H] observed by GALAH for stars in the age interval of 3--6 Gyr.  For these data an averaged shift value of $\Delta$\,[X/Fe]$_{\rm shift}\sim -0.39$\,dex and $\Delta$\,[Fe/H]$_{\rm shift}\sim +0.29$\,dex was applied. The resulted mean ($\mu$) and standard deviation ($\sigma$) of the Gaussian fit to the histograms are displayed in each plot.}
\label{gaussian}
\end{figure*}

In order to do this, we have selected the points around the solar age (3--6\,Gyr), Galactocentric radius (7--9 kpc) or metallicity, in each case. We have plotted the corresponding distribution [X/Fe] and [Fe/H] by searching for the mean value by fitting a Gaussian and obtained the mean and standard deviation from the fit. As this averaged abundances should be zero, we apply shifts values $\Delta$\,[X/Fe]$_{\rm shift}$ and $\Delta$\,[Fe/H]$_{\rm shift}$ that correspond to the averaged values obtained from the fits. We show an example of the method  in Fig.~\ref{gaussian} with data from GALAH for [O/Fe]. In this case, we applied a shift of [O/Fe]$\sim -0.39$\,dex and <[Fe/H]>$\sim +0.29$\,dex and the result is displayed in the figure. In all cases we checked that this procedure lead to the same results as calculating the mean and the standard deviation of the sample. 

\begin{figure*}
\includegraphics[width=0.8\textwidth]{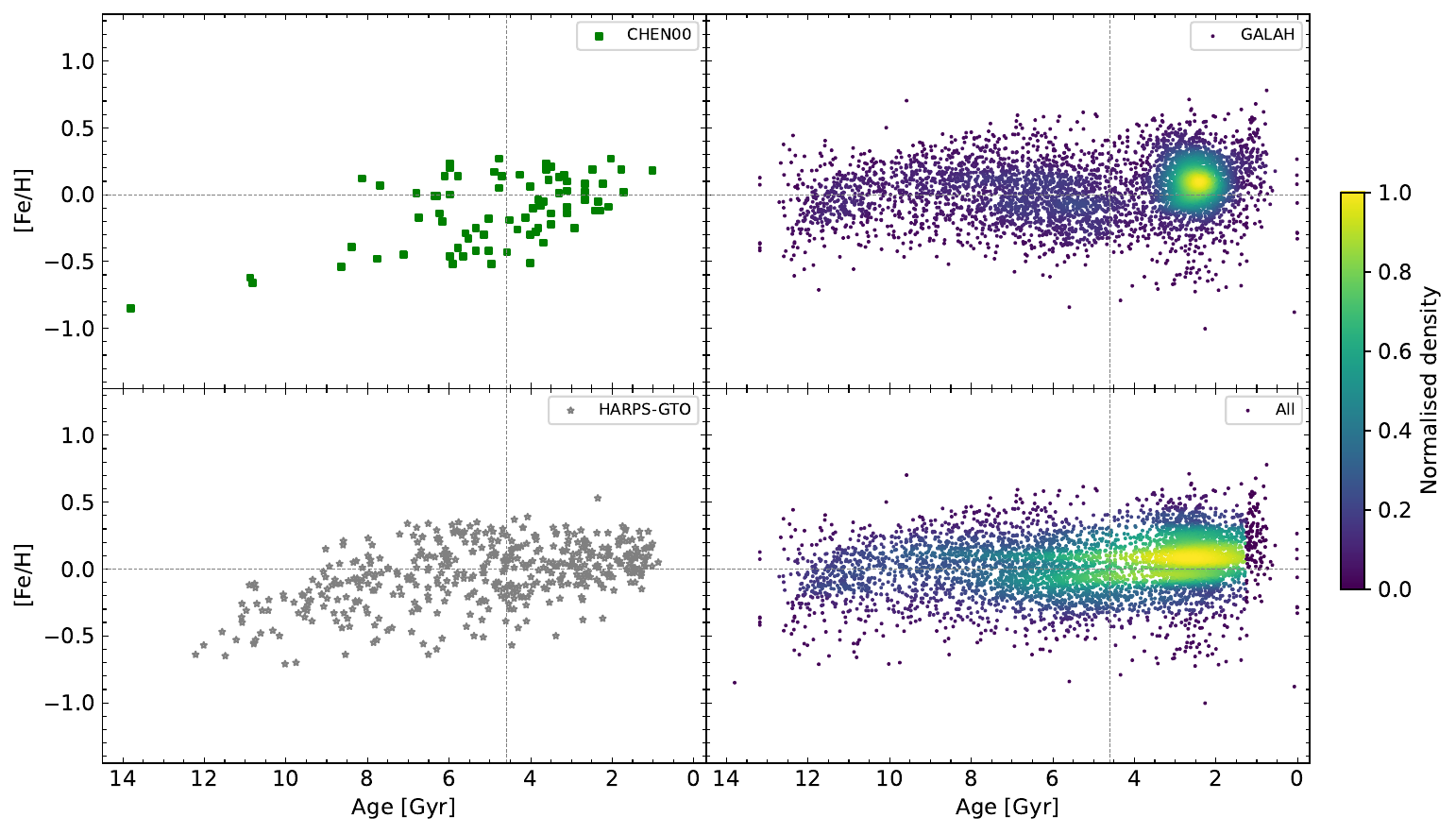}
\caption{Observational data for [Fe/H] {\sl vs.} Age from sets of different authors or surveys as labelled in panels. The bottom right panel shows all these data over-plotted. The vertical colours bars display the density scale for the respectives plots.}
\label{fig:feh_binned}
\end{figure*}

\begin{figure*}
\includegraphics[width=0.8\textwidth]{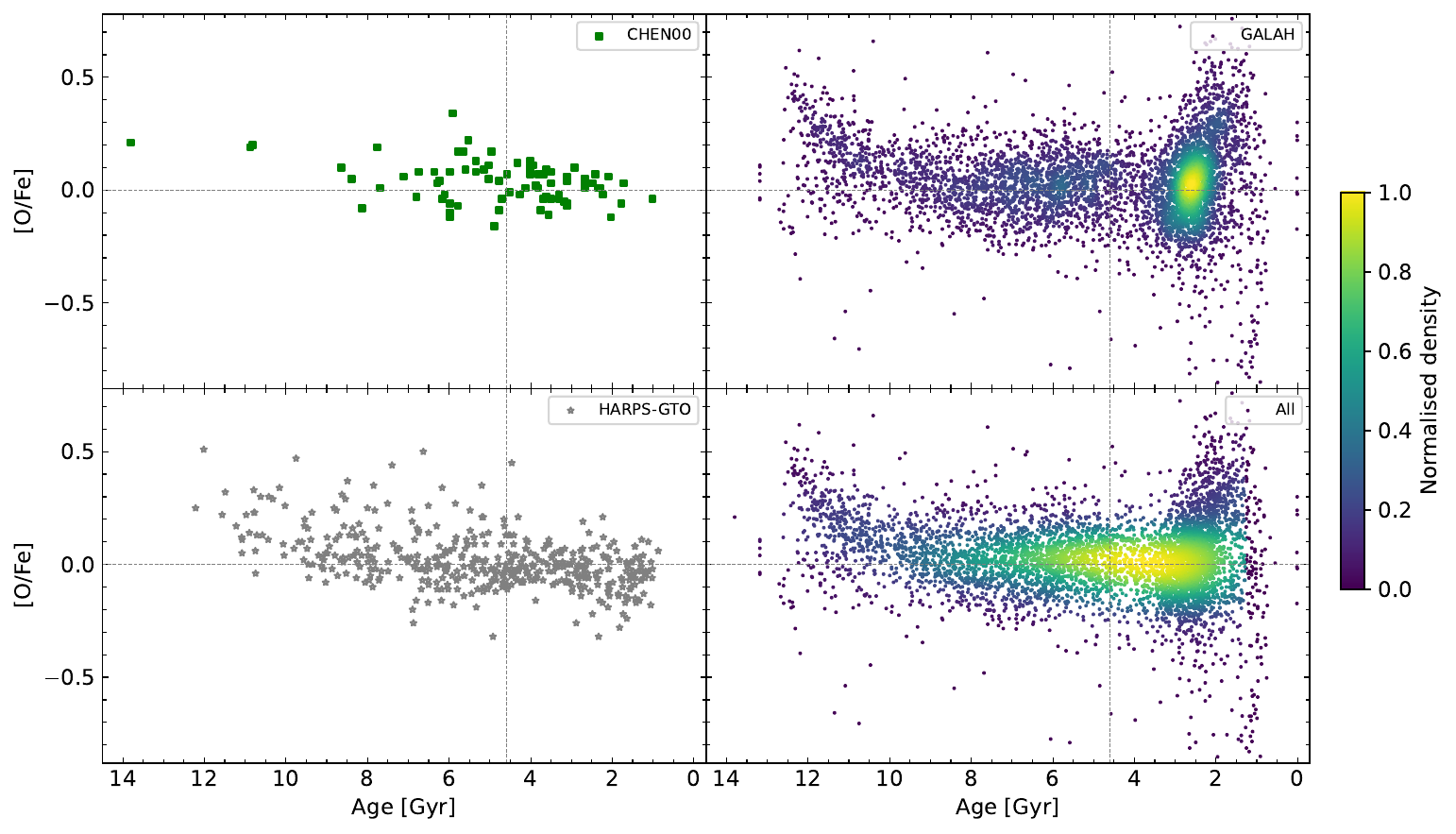} 
\caption{The same as Fig. \ref{fig:feh_binned} but for [O/Fe] {\sl vs.} Age.}
\label{fig:ofe_binned}
\end{figure*}

\begin{figure*}
\includegraphics[width=0.8\textwidth]{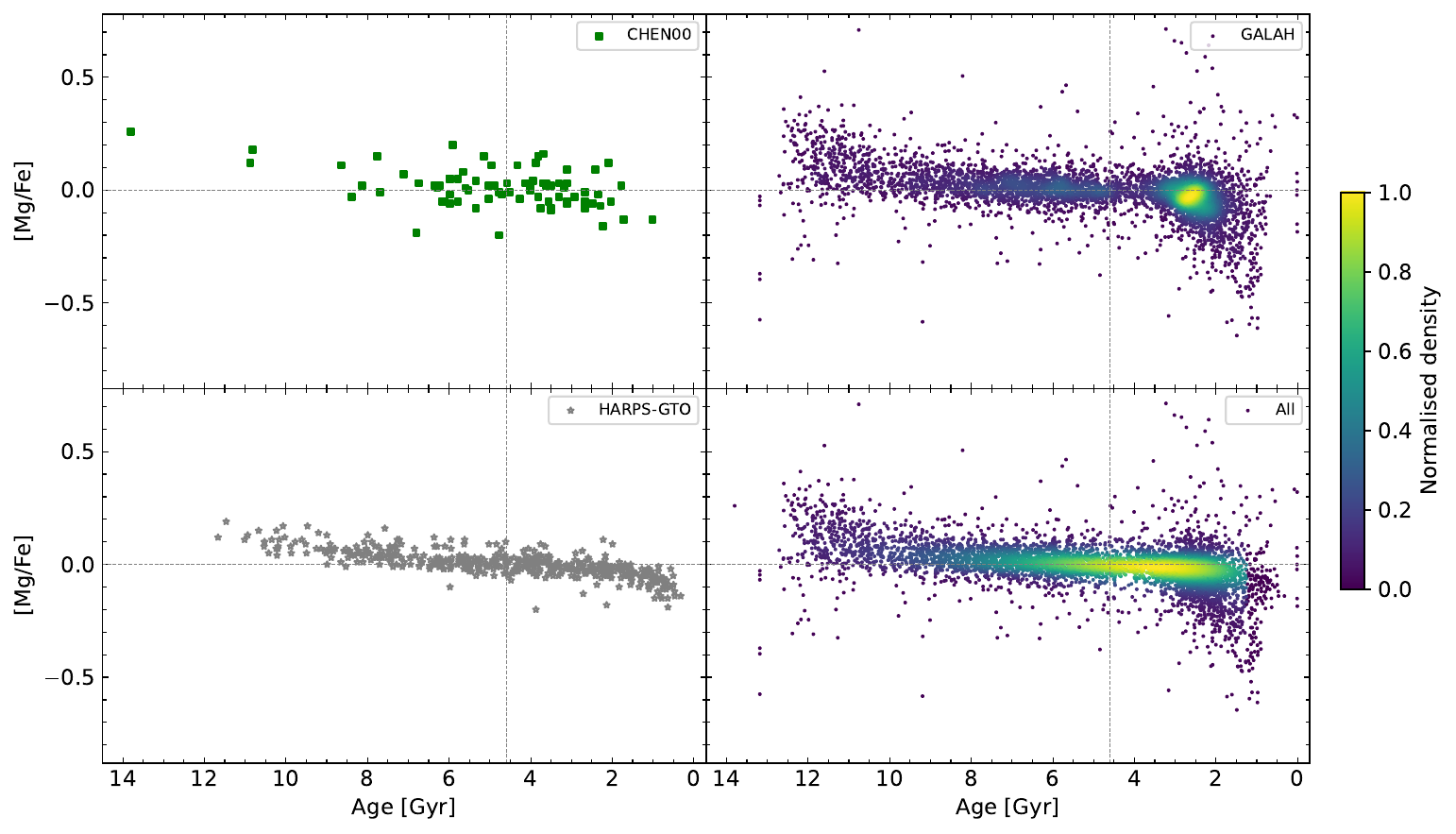}
\caption{The same as Fig. \ref{fig:feh_binned} but for [Mg/Fe] {\sl vs.} Age.}
\label{fig:mgfe_binned}
\end{figure*}

We have repeated this process for [Fe/H] and all relative abundances [X/Fe]. The resulting distributions are shown in Fig.~\ref{fig:feh_binned} for [Fe/H] as a function of age, in Figures~\ref{fig:ofe_binned}, ~\ref{fig:mgfe_binned}, ~\ref{fig:sife_binned} and  ~\ref{fig:cafe_binned}, for [X/Fe] as a function of the stellar age, and in Figs.~\ref{fig:ofe_feh_binned}, ~\ref{fig:mgfe_feh_binned} and ~\ref{fig:sife_feh_binned}, as a function of metallicity [Fe/H]. In these figures, the each dataset is represented by different symbols: green squares \citep{chen00}, grey stars (HARPS-GTO) and density (GALAH). For reference, in all of the figures we plot as dashed lines the Sun position. We have also plotted all sets together for each distribution in the last panel of each figure and these combined datasets are used in the statistical analysis given in Section \ref{Results}. 

\subsection{The age-metallicity relation}
In this section, we show the data relative to the time evolution of the Iron abundance [Fe/H], usually taken as the measurement of the global metallicity, for the solar region. 
Data are taken from different surveys as given in \citet[][ CHEN]{chen00}, \citet[][ HARPS]{dm15,dm17} and \citet[][ GALAH]{buder2021}.
Fig.~\ref{fig:feh_binned} shows 4 panels with data from CHEN, GALAH, HARPS and the bottom right panel gives the corresponding figure from all survey data sets together.

\subsection{The evolution of the $\alpha$-element abundances 
over Iron along the stellar age and [Fe/H]}

\begin{figure*}
\includegraphics[width=0.8\textwidth]{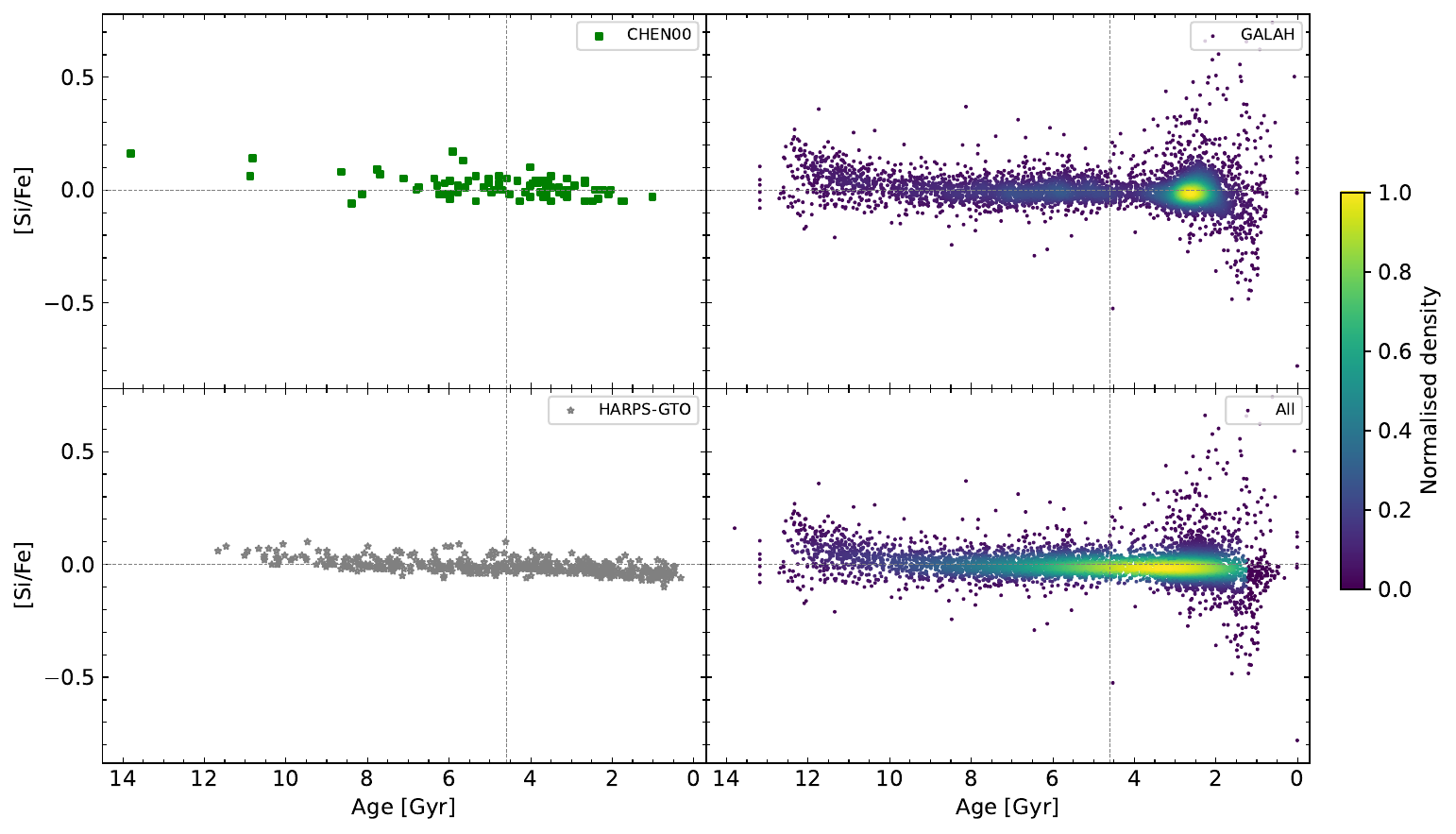}
\caption{The same as Fig. \ref{fig:feh_binned} but for [Si/Fe] {\sl vs.} Age.}
\label{fig:sife_binned}
\end{figure*}

\begin{figure*}
\includegraphics[width=0.8\textwidth]{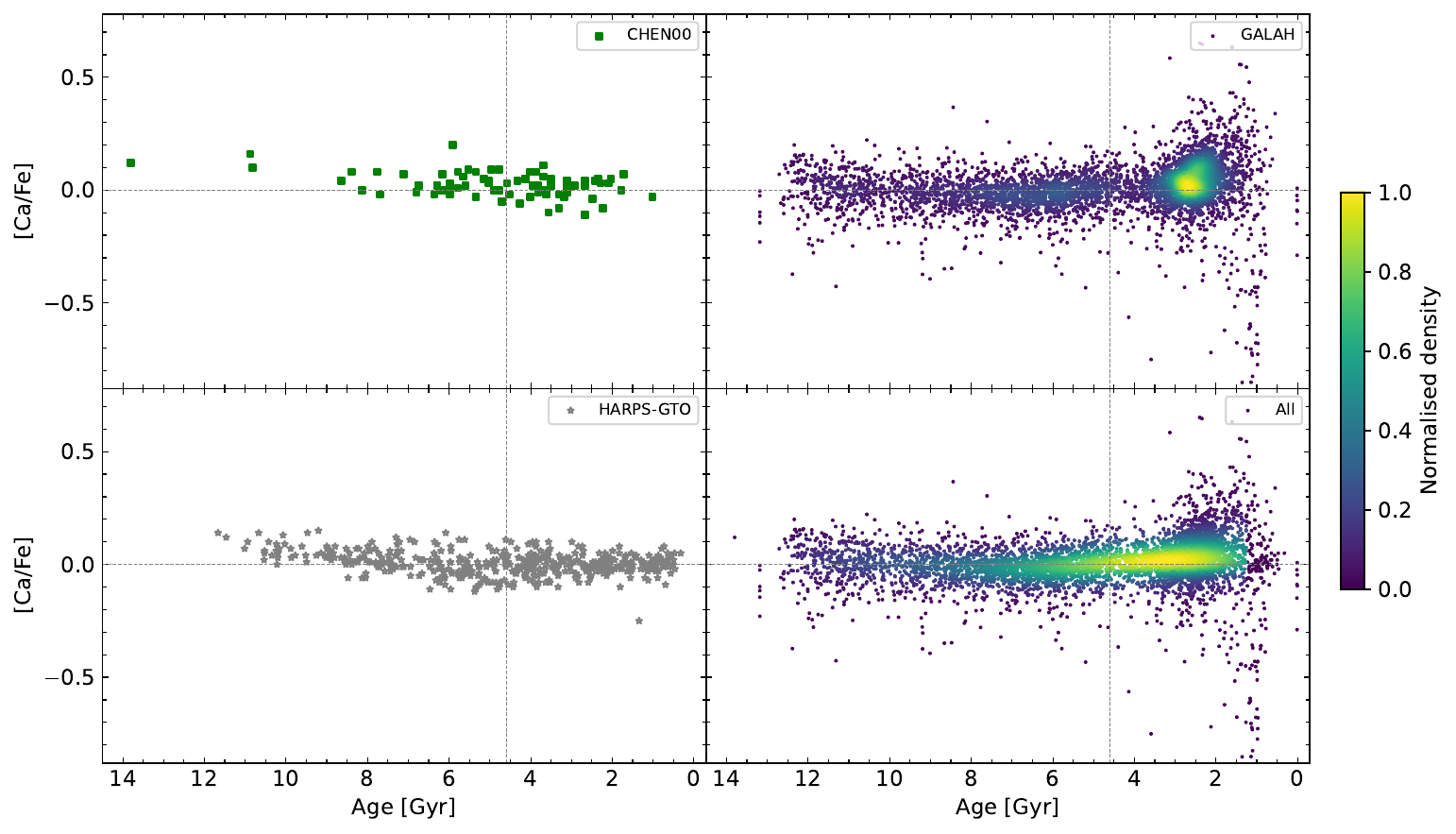}
\caption{The same as Fig. \ref{fig:feh_binned} but for [Ca/Fe] {\sl vs.} Age.}
\label{fig:cafe_binned}
\end{figure*}

In this subsection we have obtained the relative abundances [X/Fe] for O, Mg, Si and Ca as a function of the stellar age, obtained for different methods by several authors. All the data used in this work are complete in the sense that each figure has the same stars. Figures \ref{fig:ofe_binned},\ref{fig:mgfe_binned}, \ref{fig:sife_binned} and \ref{fig:cafe_binned} display the abundances ratios [O/Fe], [Mg/Fe], [Si/Fe] and [Ca/Fe] {\sl vs.} Age in Gyr, respectively. It is clear from these figures the importance of using different databases for this study, since CHEN and HARPS add stars with different ages than GALAH, increasing the densities in the plots at higher ages.

\begin{figure*}
\includegraphics[width=0.8\textwidth]{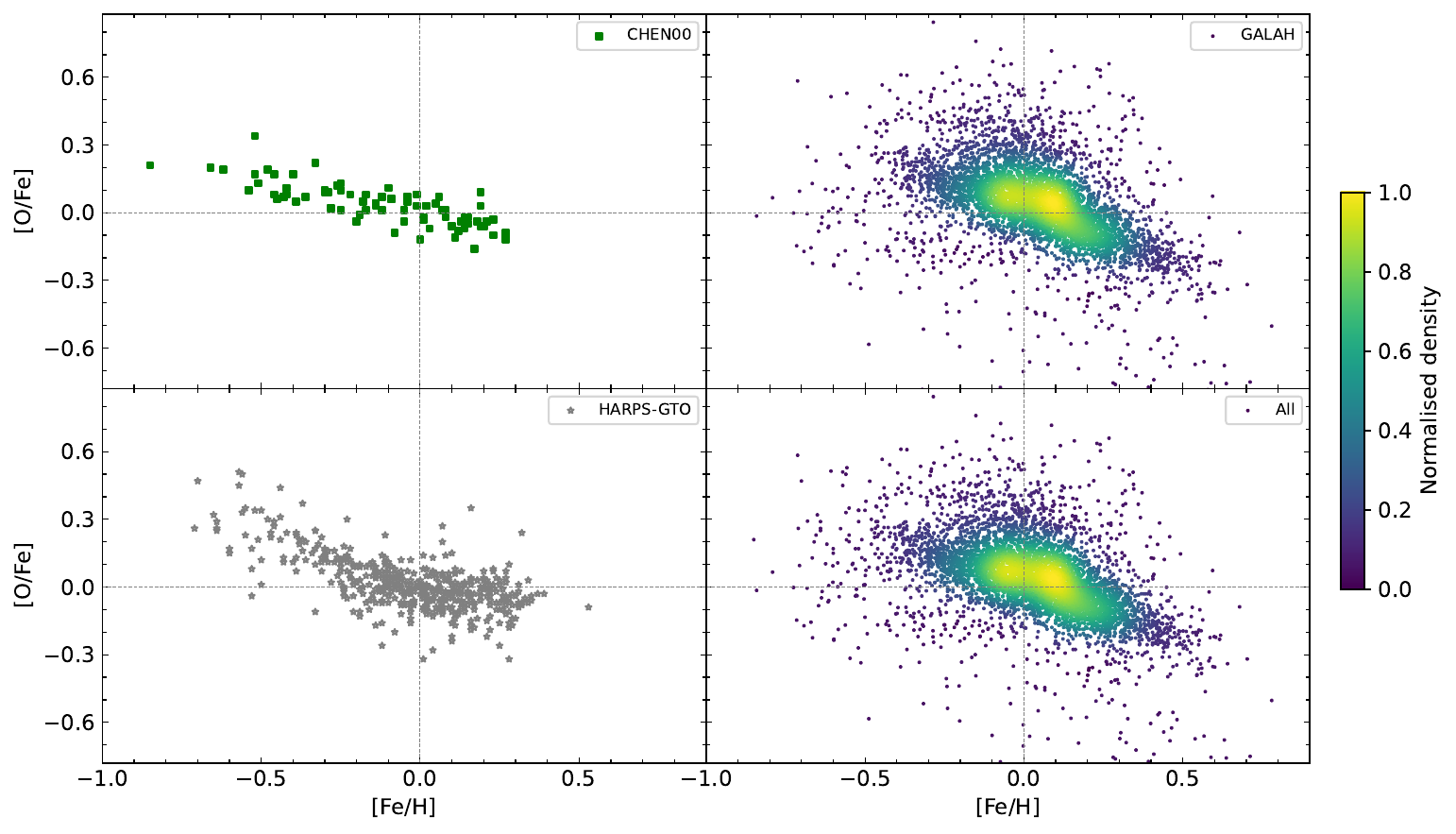}
\caption{Observational data for [O/Fe] {\sl vs} [Fe/H] from sets of different authors or surveys as labelled. The bottom right panel corresponds to the these data over-plotted.}
\label{fig:ofe_feh_binned}
\end{figure*}

\begin{figure*}
\includegraphics[width=0.8\textwidth]{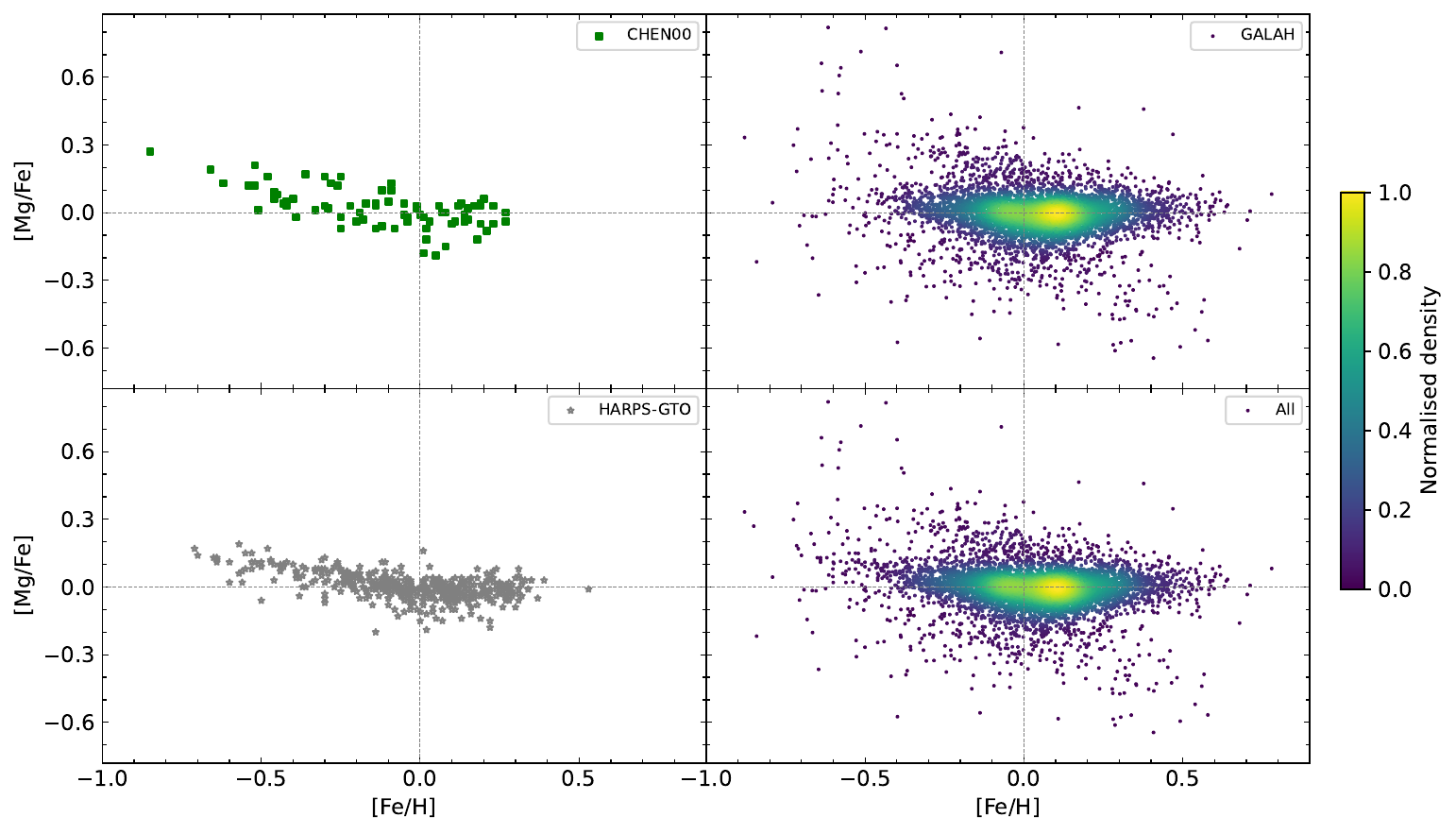}
\caption{As Fig. \ref{fig:ofe_feh_binned} but for [Mg/Fe] {\sl vs} [Fe/H].}
\label{fig:mgfe_feh_binned}
\end{figure*}

\begin{figure*}
\includegraphics[width=0.8\textwidth]{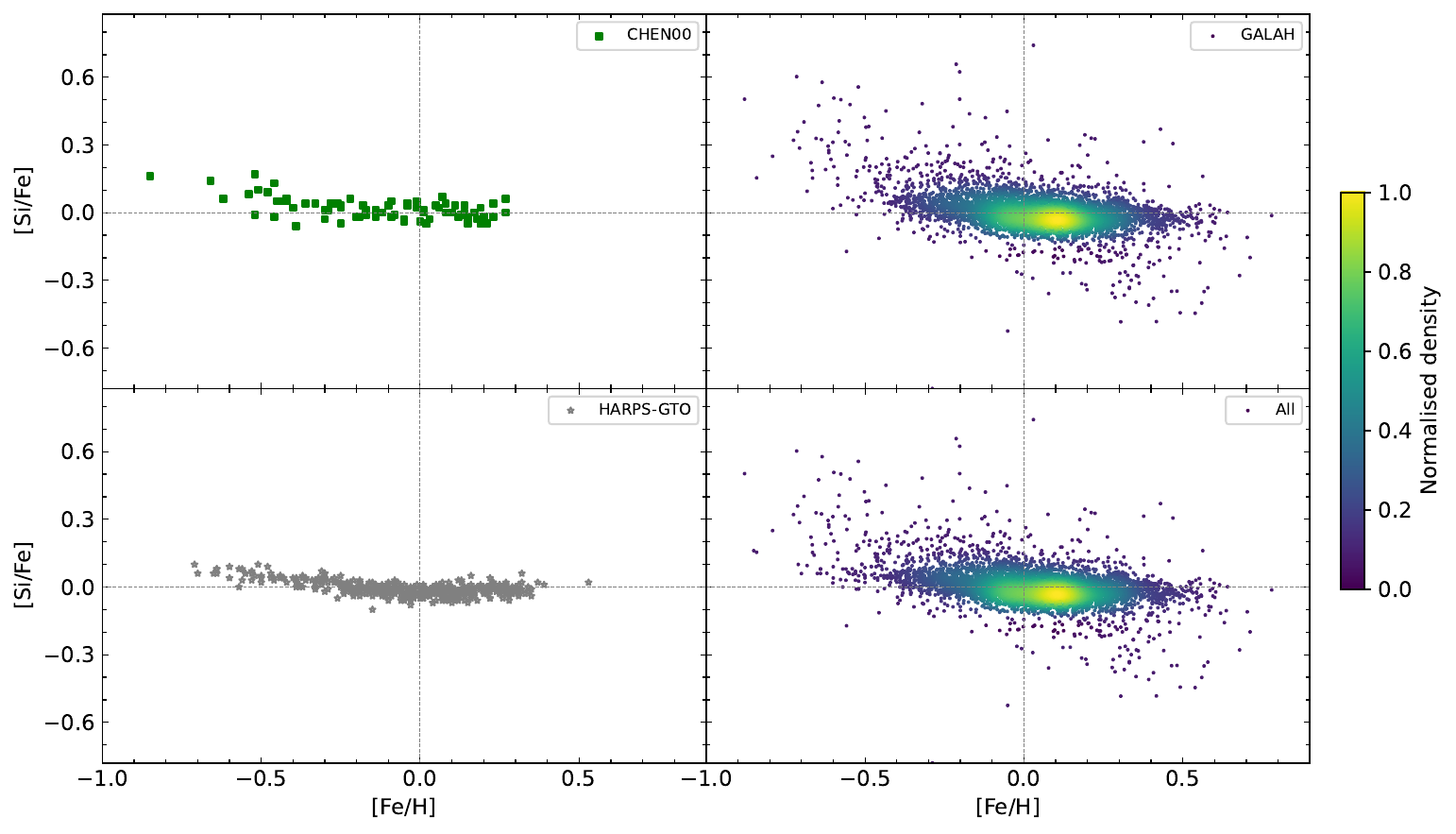}
\caption{As Fig. \ref{fig:ofe_feh_binned} but for [Si/Fe] {\sl vs} [Fe/H].}
\label{fig:sife_feh_binned}
\end{figure*}

\begin{figure*}
\includegraphics[width=0.8\textwidth]{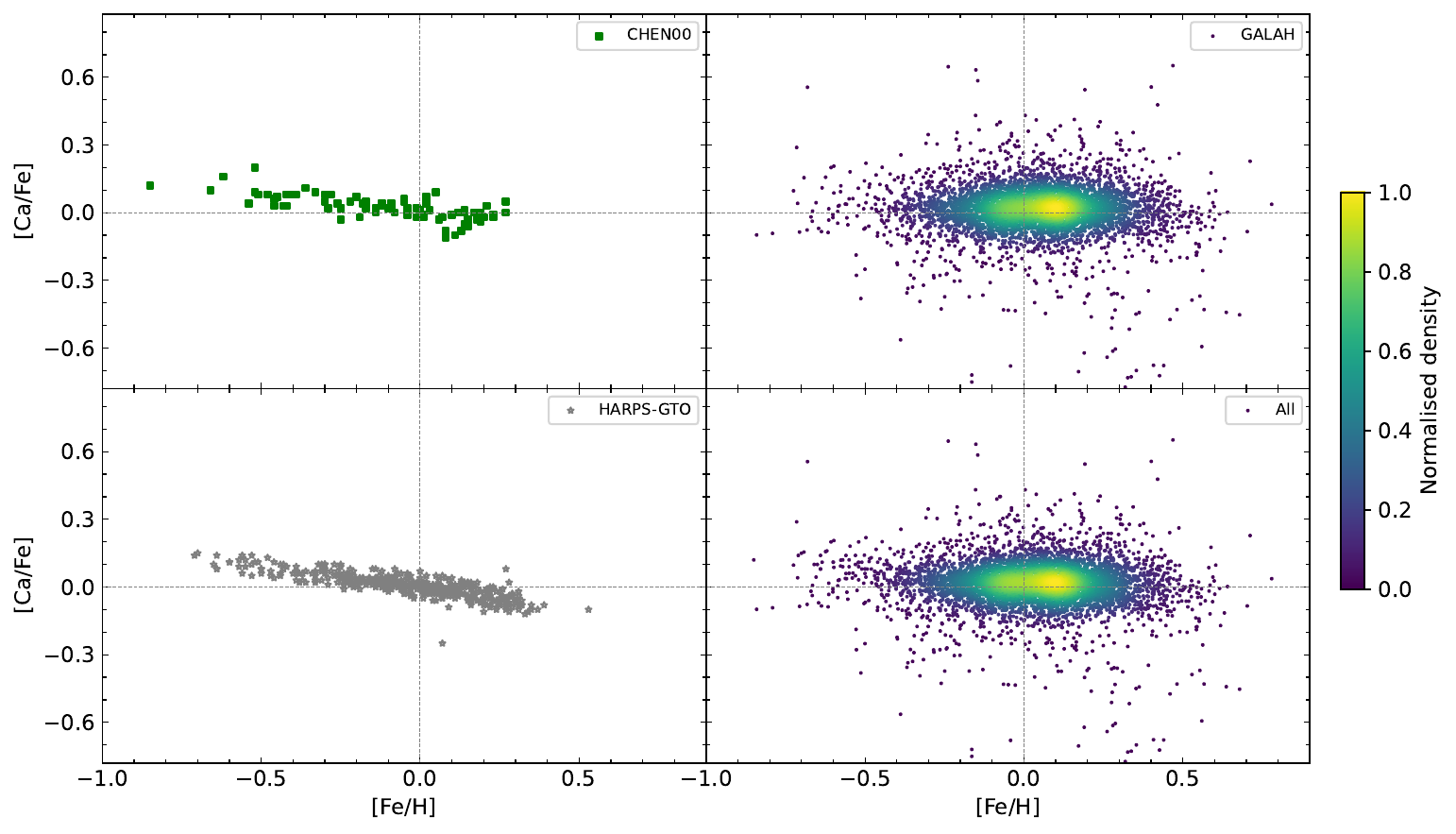}
\caption{As Fig. \ref{fig:ofe_feh_binned} but for [Ca/Fe] {\sl vs} [Fe/H].}
\label{fig:cafe_feh_binned}
\end{figure*}

Finally, we show the available data for the $\alpha$-element abundances over Iron as a function of the Iron abundance or metallicity [Fe/H],
presented in figures \ref{fig:ofe_feh_binned}, \ref{fig:mgfe_feh_binned}, \ref{fig:sife_feh_binned} and \ref{fig:cafe_feh_binned}, for the abundances ratios [O/Fe], [Mg/Fe], [Si/Fe] and [Ca/Fe] as a function of [Fe/H], respectively.

From all these figures it is possible to conform that all the databases used in this work are compatible with each other and also that they are complementary.

\section{Calibration of model for the solar region 
and Galactic disc}
\label{AppB}

\subsection{The solar vicinity}

In \citet[][ hereinafter MOL15]{mol15}, a compilation of data from different authors (see references in MOL15) for MWG was presented. A binning process resulted in a set of binned data for each type of observation. This was described in detail in the Appendix from MOL15, where a table of references of these resulting binned distributions were given. The observational data basically included the solar neighbourhood time evolution of metallicity, SFR, and CNO abundances, in addition to the present state of the disc, including the radial distributions of surface densities for stars, gas, SFR, and CNO abundances. 

In the present work, we adopted the data from references described in MOL15 for the solar vicinity time evolution of SFR. The data are binned for each Gyr with $\Delta t=\pm 0.5\,{ \rm Gyr}$. The resulting SFR values were normalised to the most recent values of the SFR in the solar region at the present time, which was obtained as SFR$_{\sun} \sim 0.266$\,\Msun\,yr$^{-1}$, at a Galactocentric distance $R = 8$\,kpc. 

The SFR and the C, N and O time evolution for the solar vicinity are plotted in Fig.~\ref{sfr} and Fig~\ref{abunt}. In these figures all the 180 models give, as expected, similar results and are in agreement with the observational data obtained in MOL15. It is evident that these quantities do not dependent on the SN Ia rates nor on the SN Ia yields --much lower than the stellar yields from massive stars/CC SNs or from intermediate stars-- shown in Fig.~5 of the main paper. Since we have updated the stellar yields for low and intermediate-mass stars, as well as massive stars, it is important to check that these new models still reproduce well the solar region observational data. We have checked that this is true by comparing the results for the 180 models –-which appear as only one—- with the data. In fact, we demonstrate that the star formation history (as the C, N and O abundances along the evolutionary time) shows good agreement compared with observational data, and effectively there is no change in the results depending on the DTD or on the SN Ia yield set.

\subsection{The Galaxy disc}

The data for radial distributions of surface densities in the MWG disc for HI, molecular gas, stellar profile and the SFR  surface densities were taken from Table A3 in MOL15. The solar stellar surface density is between 33 and 64 $M_{\sun}$\,pc$^{-2}$ from different authors, as e.g. \citet{mcmillan11, burch13, bovy13}. These values depend mostly on the scale length of the disc, $R_{\rm d}$, which is in the range [2.1--5.4]\,kpc.  
Therefore, we assume a certain uncertainty of $\sim 0.20$\,dex in the absolute values of this radial distribution. The SFR binned radial distribution was also normalised to the SFR$_{\sun}$. For what refers to the C, N and O abundances radial distributions, we have used the data of Table A4 from MOL15.
In all cases, the error bars include the statistical error that corresponds to the dispersion of the original data coming from  different authors, plus the systematical uncertainty due to the instrumental setup and to the techniques to estimate the final data from observations.

\begin{figure}
\centering
\includegraphics[width=0.32\textwidth,angle=-90]{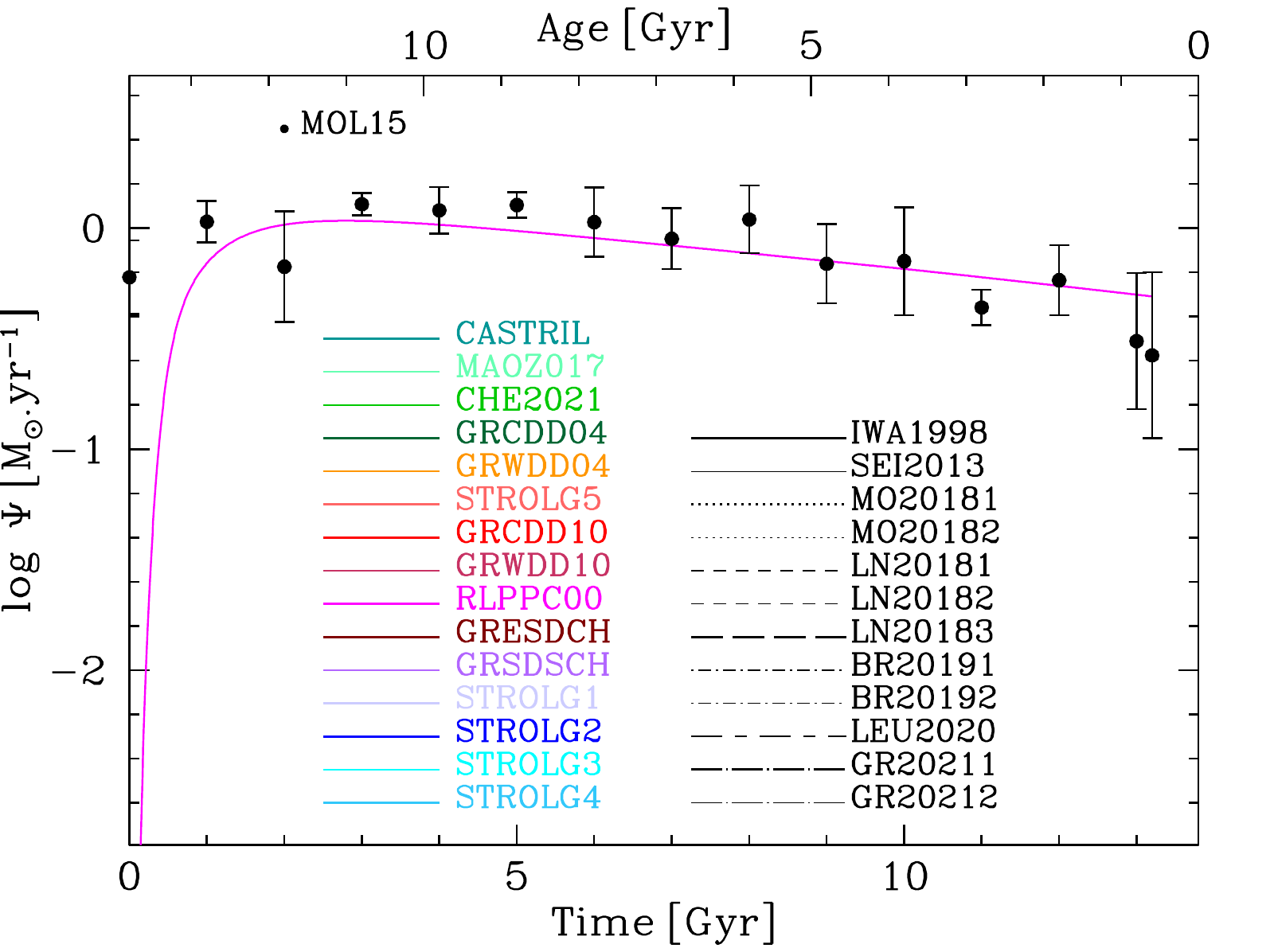}
\caption{The SFR obtained for the solar vicinity (assumed at 8\,kpc of Galactocentric distance). 
Colours and line styles indicate different models (perfectly overlapping) as labelled in Tables \ref{tab:dtdmodels} and \ref{tab:sniayields} from the main paper.
Black full dots represent the observational data compiled by \citet[][ Appendix]{mol15}.} 
\label{sfr}
\end{figure}

\begin{figure}
\centering
\includegraphics[width=0.4\textwidth]{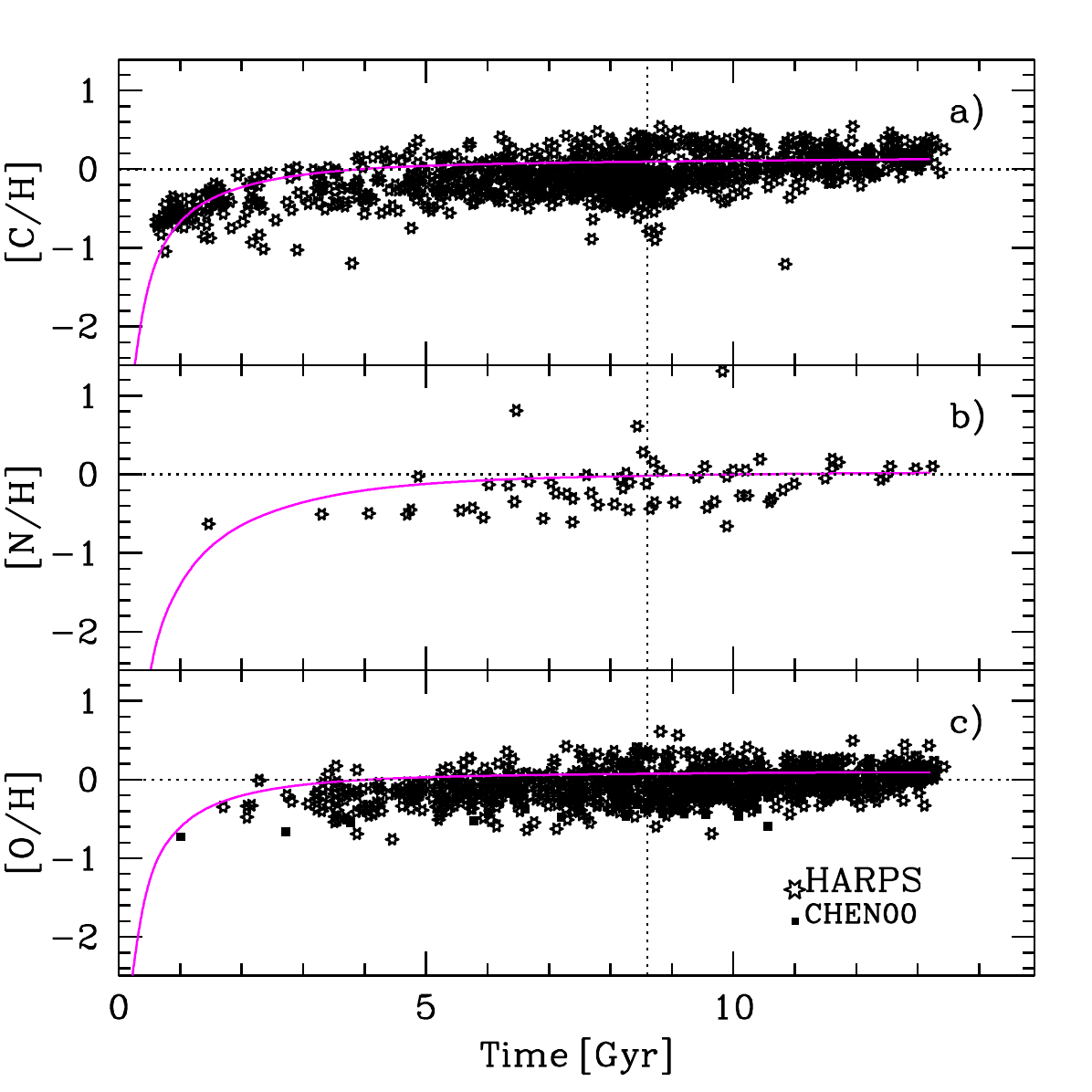}
\caption{Time evolution of a) C; b) N; and c) O abundances, as [X/H] where the solar abundances are taken from \citet{Lodders2019} for the complete set of models.
Stellar abundance data are from \citet{chen00} and from the HARPS-GTO collaboration (see text for references). The dotted lines indicate the Sun position at its birth's time, for reference.
Different models are plotted as (overlapping) lines following the same style and colour scheme as in Fig.~\ref{sfr} although only a line is apparent since all models fall in the same locus.
}
\label{abunt}
\end{figure}

\begin{figure}
\centering
\includegraphics[width=0.4\textwidth, angle=0]{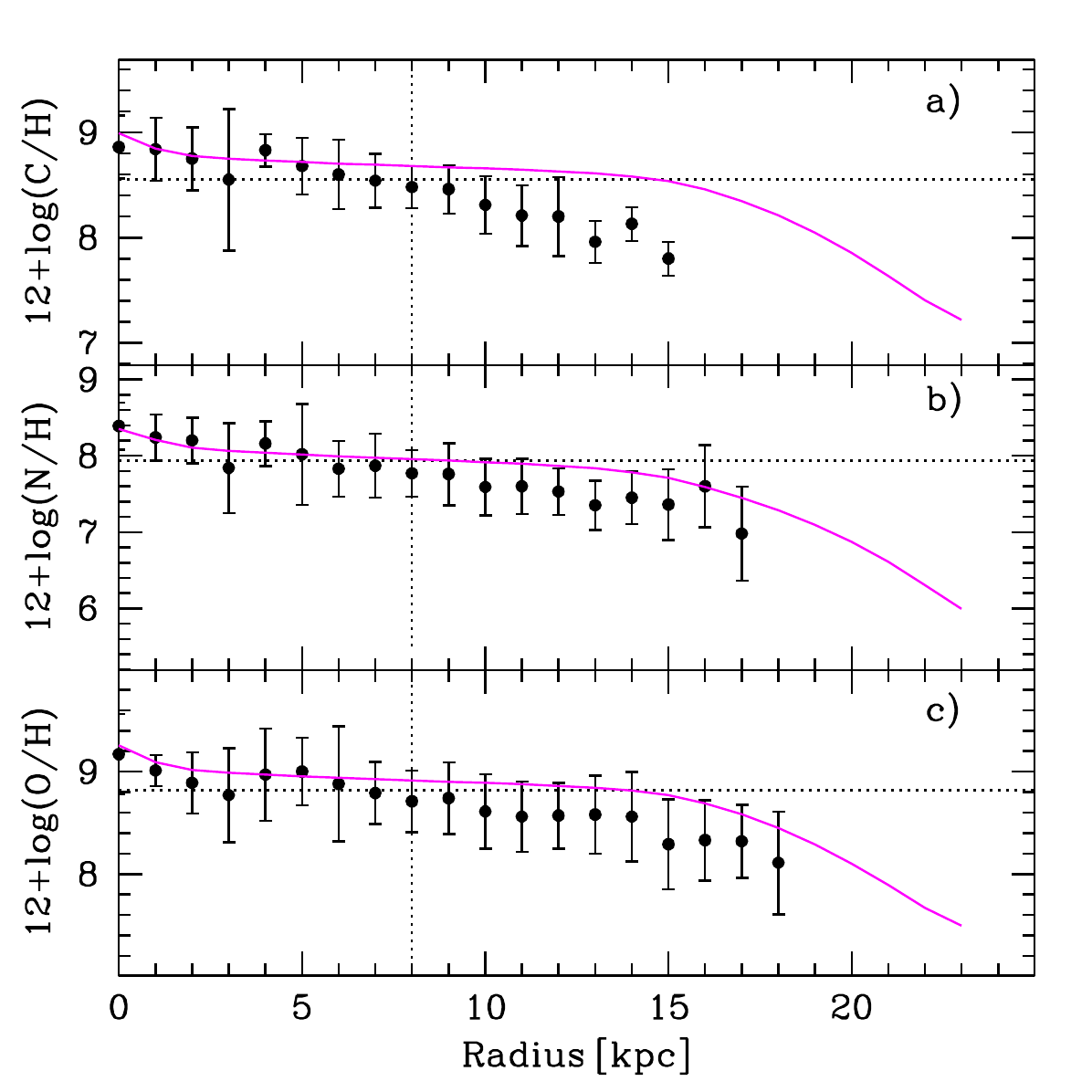}
\caption{The radial distribution of elemental abundances of a) C; b) N; and c) O as $12+\log(\rm{X}/\rm{H})$, for the complete set of models. The black full dots are observational data taken from \citet[][ Appendix]{mol15}. The dotted lines indicate the Sun abundance at its position $R=8.5$\,kpc, for reference.
Different models are plotted as (overlapping) lines following the same style and colour scheme as in Fig.~\ref{sfr} although only a line is apparent since all models fall in the same locus.
}
\label{ab-R}
\end{figure}

\begin{figure}
\centering
\includegraphics[width=0.35\textwidth,angle=-90]{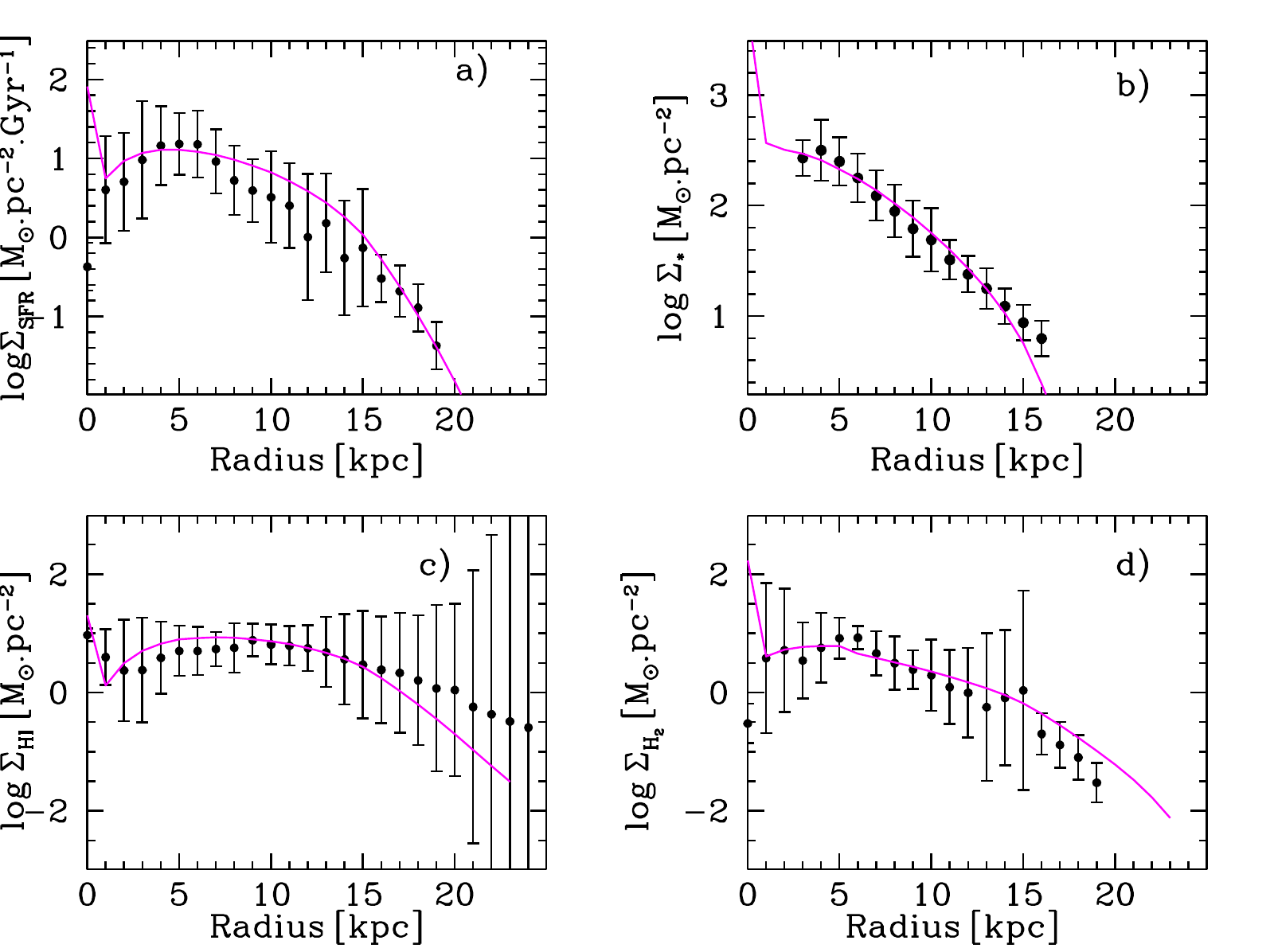}
\caption{Radial densities distributions for the Galactic disc for the complete set of models. Panels correspond to: a) SFR; b) stellar mass; c) diffuse gas and d) molecular gas, all in units of M$_{\sun}\,{\rm pc}^{-2}$, except for the SFR, which is in M$_{\sun}\,{\rm pc}^{-2}\,{\rm Gyr}^{-1}$). Black full dots are binned data from \citet[][ Appendix]{mol15}, while theoretical predictions are apparently shown with only a line because all models perfectly overlap in all panels.}
\label{disco}
\end{figure}

Next figures refer to the radial distribution of different quantities throughout the Galactic disc. We show the radial distributions of C, N and O in Fig~\ref{ab-R}. Fig.~\ref{disco} displays the surface density radial distributions of the SFR, the stellar profile, and both gas phases, diffuse and molecular. In all figures the corresponding data are those compiled in MOL15 with their error bars, which indicate the dispersion of the original data from different authors. In both figures, the results of all models are identical and the lines overlap each other. In all cases, the models still present satisfactory results in agreement with the observational data, taking into account the data spread and uncertainties. 

We did not expect large differences between models for C, N and O, which basically come from stars of different masses, although there are small quantities coming from SN Ia. As explained above, we have included Fig.~\ref{abunt}, and \ref{sfr}  as Fig.~\ref{ab-R} and \ref{disco}, to demonstrate that using different sets for stellar yields than MOL15 and MOL17 give essentially the same evolution for the solar region and the Galactic disc, that is, the new model is well calibrated compared with the data for these regions.

\bsp	% typesetting comment
\label{lastpage}
\end{document}